\documentclass[notitlepage,prd,twocolumn,showpacs,preprintnumbers,nofootinbib,superscriptaddress,floatfix,tightenlines]{revtex4-1}
\usepackage{mathrsfs}
\usepackage{amssymb}
\usepackage{amsmath}
\usepackage{graphicx}
\usepackage{amsfonts,amsbsy}
\usepackage{nicefrac}
\usepackage{xcolor}
\usepackage[breaklinks,colorlinks,urlcolor=blue,citecolor=citcolor,linkcolor=lcolor]{hyperref}
\usepackage{longtable}

\definecolor{lcolor}{rgb}{0.5,0,0}
\definecolor{citcolor}{rgb}{0,0.3,0.0}
\definecolor{ao(english)}{rgb}{0.0, 0.5, 0.0}
\usepackage{comment}

\usepackage{graphicx}
\usepackage{wrapfig}

\usepackage{physics}
\usepackage{hyperref}
\usepackage{multirow}
\usepackage{array}
\usepackage{empheq}

\newcommand{\mbf}{\mathbf}

\newcommand{\mrm}{\mathrm}

\newcommand{\wplas}{\omega_{\mrm{pl}}}

\newcommand{\fig}{Fig.~}
\newcommand{\figs}{Figs.~}
\newcommand{\eq}{Eq.~}
\newcommand{\eqs}{Eqs.~}
\newcommand{\se}{Sec.~}
\newcommand{\ses}{Secs.~}
\newcommand{\re}{Ref.~}
\newcommand{\res}{Refs.~}
\newcommand{\app}{Appendix~}
\newcommand{\apps}{Appendices~}

\newcommand{\tab}{Table~}

\newcommand{\pToFigs}{Images}

\usepackage{xargs}
\usepackage[colorinlistoftodos,prependcaption,textsize=footnotesize]{todonotes}
\newcommandx{\LB}[2][1=]{\todo[linecolor=orange,backgroundcolor=orange!25,bordercolor=orange,#1]{LB: #2}}
\newcommandx{\KB}[2][1=]{\todo[linecolor=purple,backgroundcolor=purple!25,bordercolor=purple,#1]{KB: #2}}
\newcommandx{\PH}[2][1=]{\todo[linecolor=yellow,backgroundcolor=yellow!25,bordercolor=yellow,#1]{PH: #2}}
\newcommandx{\TODO}[2][1=]{\todo[linecolor=red,backgroundcolor=red!25,bordercolor=red,#1]{TODO: #2}}

\begin{document}

\title{Heavy-quark diffusion in 2+1D and Glasma-like plasmas: evidence of a transport peak}

\author{Liane Backfried} 
\author{Kirill Boguslavski} 
\author{Paul Hotzy} 
\affiliation{Institute for Theoretical Physics, TU Wien, Wiedner Hauptstraße 8-10, 1040 Vienna, Austria}

\begin{abstract}
    We examine how plasma excitations affect the heavy-quark diffusion coefficient in 2+1 dimensional and Glasma-like plasmas, akin to the pre-equilibrium matter describing relativistic heavy-ion collisions at early times. We find that the transport coefficient $2\kappa$ transverse to and $\kappa_z$ along the beam direction display a qualitatively different evolution. We attribute this to the underlying excitation spectra of these systems, thus providing evidence for the existence of non-perturbative properties in the spectrum. This is accomplished by first reconstructing the diffusion coefficients using gauge-fixed correlation functions, which accurately reproduces the time evolution of $2\kappa$ and $\kappa_z$. We then modify these excitation spectra to study the impact of their non-perturbative features on the transport coefficients. In particular, we find evidence for a novel transport peak in the low-frequency spectrum that is crucial for heavy-quark diffusion. We also demonstrate that gluonic excitations are broad while scalar excitations associated with $\kappa_z$ are narrow and strongly enhanced at low momenta. Our findings indicate that the large values and dynamical properties of transport coefficients in the Glasma could originate from genuinely non-perturbative features in the spectrum.
\end{abstract}

\maketitle


\section{Introduction} \label{sec:introduction}
    In the initial stages of heavy-ion collisions, an exotic state of QCD matter, the quark-gluon plasma (QGP), is formed \cite{Busza:2018rrf}. Despite intense studies, understanding the properties in different stages of its evolution, particularly the pre-equilibrium matter, remains a significant challenge in theoretical physics. These early stages may exhibit unique and non-perturbative properties that are not easily accessible through direct observation or computational techniques. As a result, open questions persist regarding the behavior of the QGP during its approach toward a hydrodynamic evolution. 

    One promising method to probe the pre-equilibrium and non-perturbative features of the QGP is through the study of transport phenomena using hard probes. By comparing theoretical predictions with experimental observables, particularly in the context of heavy-quark diffusion and quarkonium suppression, one can gain valuable insights into the underlying physics of the QGP \cite{He:2022ywp, Apolinario:2022vzg}. Heavy quarks, due to their large masses, serve as excellent probes for these phenomena, with the heavy-quark (momentum) diffusion coefficient $\kappa$ playing a critical role in characterizing their interactions with the medium \cite{Moore:2004tg, Rapp:2018qla}. This coefficient enters the diffusion process for heavy quarks and, therefore, plays a crucial role in the study of heavy-quark flow and spectra \cite{Moore:2004tg, Rapp:2018qla, Xu:2018gux, Cao:2018ews}. Furthermore, quarkonia can be studied in an open quantum systems approach that may require $\kappa$ as an input parameter \cite{Rothkopf:2019ipj, Andronic:2024oxz, Brambilla:2016wgg, Brambilla:2020qwo, Brambilla:2023hkw}. 

    While significant efforts have been made in calculating $\kappa$ in thermal equilibrium using perturbation theory \cite{Moore:2004tg, Caron-Huot:2007rwy, Caron-Huot:2008dyw} and lattice QCD calculations \cite{Brambilla:2020siz, Banerjee:2022gen, Brambilla:2022xbd, Altenkort:2023oms, Altenkort:2023eav, Laine:2009dd}, recent studies have begun to explore $\kappa$ also during the pre-equilibrium stages preceding the (thermalized) QGP. In particular, the pre-equilibrium evolution \cite{Berges:2020fwq} can be divided into the Glasma phase consisting of highly occupied QCD matter far from equilibrium \cite{Gelis:2012ri} that can be simulated using a classical-statistical approach \cite{Krasnitz:1998ns, Lappi:2003bi, Berges:2013fga}, and the subsequent kinetic phase \cite{Baier:2000sb, Kurkela:2015qoa} that is dominated by scatterings of quasi-particle excitations \cite{Boguslavski:2018beu, Boguslavski:2021kdd}. During these stages, the heavy-quark diffusion coefficient in the Glasma exhibits large anisotropic values in different directions \cite{Carrington:2020sww, Boguslavski:2020tqz, Khowal:2021zoo, Avramescu:2023qvv, Das:2015aga, Mrowczynski:2017kso, Ipp:2020mjc, Pandey:2023dzz}, and the kinetic regime connects these with the later hydrodynamic evolution \cite{Boguslavski:2023fdm, Du:2023izb, Boguslavski:2023jvg, Romatschke:2006bb}. Each of the regimes is expected to affect the nuclear suppression factor $R_{AA}$ and elliptic flow of heavy quarks \cite{Das:2017dsh, Sun:2019fud, Prakash:2023wbs}.

    Some important features of $\kappa$ in the Glasma can be understood already in non-expanding non-Abelian systems. For instance, oscillations in the evolution of $\kappa$ visible in (isotropic) 3+1 dimensional highly occupied plasmas have been connected to an enhancement of quasi-particle excitations in the pre-equilibrium matter \cite{Boguslavski:2020tqz}. However, in the Glasma, the time evolution of the heavy-quark diffusion coefficients transverse to and in the beam direction $2\kappa$ and $\kappa_z$ is qualitatively different, where $\kappa_z$ can even become negative \cite{Avramescu:2023qvv}, which is not captured by such 3+1D models. Instead, it proves suitable to describe the Glasma in terms of an effectively 2+1D system due to its (approximate) boost invariance. However, for such a dimensionally reduced system, hard thermal loop (HTL) perturbation theory \cite{Braaten:1989mz, Blaizot:2001nr} breaks down, as soft momenta become equally important in the evaluation of loops and the underlying assumption of a scale separation between hard and soft modes is violated \cite{Boguslavski:2019fsb}. This highlights the limitations of traditional approaches and the need for alternative methods to gain more insight.

    Recent observations in different 2+1 dimensional Glasma-like systems have revealed excitations that exhibit strong non-perturbative behavior \cite{Boguslavski:2021buh}. In particular, correlation functions in temporal and (spatial) Coulomb gauge display broad gluonic peaks, an enhancement of low-momentum `scalar' excitations, and, most intriguingly, an excitation at zero frequency that we will associate with a new transport peak. These findings suggest that instead of relying on the HTL approximation, one can use these excitations to reconstruct the time evolution of the gauge invariant $\kappa$. Manipulating the excitation spectra enables the study of how their features affect the transport properties of the plasma, which is analogous to the strategy that was successfully applied to non-Abelian plasmas in 3+1 dimensions \cite{Boguslavski:2020tqz}. This approach not only helps to elucidate the difference between $2\kappa$ and $\kappa_z$ but also provides physical evidence for the presence of non-perturbative features in the gauge-fixed correlators.

    In this paper, we follow this strategy and study the heavy-quark diffusion coefficient in different 2+1D Glasma-like systems. These are highly occupied plasmas in 2+1 dimensions without or with an additional scalar field, which mainly differ from the original Glasma by dropping the (longitudinal) Bjorken expansion of the metric \cite{Boguslavski:2019fsb, Boguslavski:2021buh}. We compute the evolution of $2\kappa$ and $\kappa_z$ directly and reconstruct them from frequency-dependent and gauge-fixed correlation functions. Then we conduct a detailed study of the microscopic excitations and identify a new transport peak at low momenta. By modifying the correlators, we find that it, together with the widths and amplitudes of gluonic and scalar excitations, has a crucial impact on the transport coefficients. In particular, we demonstrate how this non-perturbative structure of the excitation spectrum leads to qualitatively different properties of $2\kappa$ and $\kappa_z$ and argue that this yields a possible explanation for the large values of the heavy-quark diffusion coefficient in the Glasma evolution.

    This work is structured as follows: In \se\ref{sec:theory}, we review the theoretical background, our setup, and relevant quantities computed in this paper. We introduce the heavy-quark diffusion coefficient $\kappa(t, \Delta t)$ in \se\ref{sec:hq_diff_in_2d} and discuss its properties such as its self-similarity and (universal) time-$\Delta t$ evolution. In \se\ref{sec:spectral_reconst}, we reconstruct this observable using correlation functions in the frequency domain. In \se\ref{sec:properties_reconst}, we systematically manipulate the excitation spectrum and study the impact on the heavy-quark diffusion coefficients. We discuss the implications of our findings for heavy-ion collisions and especially for the Glasma in \se\ref{sec:Glasma}, before concluding in \se\ref{sec:conclu}. Further details of the HTL formalism used in the spectral reconstruction procedure and our results on self-similar properties can be found in \apps \ref{app:HTL} and \ref{app:self_sim_parameters}. The data for the computations performed in this work, including scripts for the reproduction of the figures, are available in \cite{backfried_2024_14039178}.


\section{Theoretical background} \label{sec:theory}
    In this section, we introduce the gauge theories considered in this work and describe the simulation method, correlation functions, initial conditions, as well as the emerging self-similar state in which we will study heavy-quark diffusion further below.

\subsection{Theories and simulation method} \label{subsec:intro_theories}
   Motivated by the Glasma model during the initial stages in heavy-ion collisions, we study two non-Abelian SU($N_c$) gauge theories with Yang-Mills action
   \begin{equation} \label{eq: YMaction}
        S_\mrm{YM}[A] = - \frac{1}{4} \int \mathrm{d}^{d+1}x F^{\mu\nu}_a F_{\mu\nu}^a,
    \end{equation}
    where $A^a_\mu$ are gauge fields and $F_{\mu\nu}^a = \partial_\mu A^a_\nu - \partial_\nu A^a_\mu + g f^{abc} A^b_\mu A^c_\nu$ is the field strength tensor in adjoint representation. Repeated color indices $a = 1, \dots, d_A$ and Lorentz indices $\mu, \nu$ = 0, \dots, $d$ imply summation over them, where $d_A = N_c^2-1$ is the number of gauge degrees of freedom. Both theories are in (effectively) $d=2$ spatial dimensions and we use $N_c = 2$ in our simulations. 
   
    The first theory (henceforth referred to as \emph{`2D'}) is a 2+1 dimensional Yang-Mills theory with classical action
    \begin{equation} \label{eq: YMaction_2D}
        S_\mrm{YM}^\mrm{2D}[A] = \left. S_\mrm{YM}[A]\right|_{d=2}\,.
    \end{equation}
    The second theory is derived from a Yang-Mills theory in originally $d=3$ spatial dimensions where the fields do not exhibit any dependence on the third coordinate $z$. This reduces it to an effectively 2+1 dimensional gauge theory, which can be written in terms of the former 2D theory for gauge bosons and an additional adjoint scalar field $A^a_z$. Consequently, its classical action reads
    \begin{equation} \label{eq: 2Dscaction}
        S_{\mathrm{YM}}^{\mathrm{2D+sc}}[A] = S_{\mathrm{YM}}^{\mathrm{2D}}[A] + S_{\mathrm{sc}}^{\mathrm{2D}}[A_z]
    \end{equation}
    and is composed of $S_\mrm{YM}^\mrm{2D}$ and the scalar field action
    \begin{equation} \label{eq: phi_action}
        S_{\mathrm{sc}}^\mrm{2D}[A_z] = - \frac{1}{2} \int \mathrm{d}^{2+1}x (D^{ab}_j A_z^b)(D_{ac}^j A_z^c).
    \end{equation}
    Here, the summation runs over $j = 1, 2$ and the covariant derivative reduces to $D^{ab}_j = \delta^{ab}\partial_j - g f^{abc} A^c_j$. We will refer to this model as \emph{`2D+sc'} as well as \emph{`Glasma-like'} as it can be interpreted as a simplified model of the Glasma in heavy-ion collisions \cite{Krasnitz:1998ns, Lappi:2003bi, Gelis:2012ri}. Our model describes a gluonic plasma where the color fields exhibit independence in the $z$ direction. In contrast to the Glasma that is commonly described in a Bjorken expanding spacetime where boost invariance in rapidity direction is assumed, we treat the gluonic plasma in Minkowski spacetime.

    This early stage of the pre-equilibrium matter in heavy-ion collisions is typically modeled with large gauge fields, which we also use for our systems. This justifies using the classical-statistical simulation method in a weak-coupling scenario $g^2 / Q \ll 1$ to study the pre-equilibrium dynamics of such systems \cite{Aarts:2001yn, Smit:2002yg, Kurkela:2012hp, Berges:2013lsa}. The quantity $Q$ is a characteristic (saturation) momentum scale set initially.

    The theory is discretized on a spatial $N_s^2$ lattice with lattice spacing $a_s$. The real-time dynamics follows from solving a set of discretized classical Hamiltonian equations of motion in a leapfrog scheme that requires a small time step $a_t \ll a_s$. The equations are solved for gauge-covariant link fields $U_j(t, \textbf{x})$ and chromo-electric fields $E^j_a = \partial_t A^a_j$ in temporal gauge $A_0 = 0$ where the Gauss law is restored after initialization using an algorithm from \cite{Moore:1996qs}. The link variables are defined via the gauge fields by $U_j(t, \textbf{x}) \approx \mathrm{exp}(iga_sA_j(t, \textbf{x}))$ and ensure gauge invariance in the formulation of the equations of motion.%
    \footnote{
        The approach and implemented algorithm are described in more detail in \re \cite{Boguslavski:2018beu} and references therein and have been applied to the considered theories in \res \cite{Boguslavski:2019fsb, Boguslavski:2021buh}. In the 2+1D model, we use $j = 1, 2$, whereas the same scheme applies for the Glasma-like system with $j = 1, 2, z$ but small longitudinal spacing $a_z \to 0$.
    }

    Besides the temporal gauge, there is a residual spatial gauge freedom in the considered theories. We choose to fix to Coulomb-like gauge $\partial_j A_j = 0 \vert_t$ at time $t$ that is commonly used in classical-statistical gauge simulations \cite{Lappi:2003bi, Kurkela:2012hp, Berges:2007re, Berges:2013fga, Berges:2013eia}. It has the advantage that it suppresses the longitudinal polarization in equal-time correlation functions at sufficiently high momenta above the Debye-mass scale. Therefore, the remaining transversely polarized statistical correlation function (see \se \ref{subsec:corr_funcs}) suffices to define the quasiparticle distribution, i.e., occupation numbers $f(t, p) = \langle E E \rangle_T(t, t, p)\,/\,(d_A\, p)$.

\subsection{Spectral and statistical correlation functions} \label{subsec:corr_funcs}
    For our study, the measurement of equal- and unequal-time correlation functions is of central interest. They hold information about the excitation spectrum and the quasiparticle character of the medium and will be used in the computation and reconstruction of the heavy-quark diffusion coefficient $\kappa(t, \Delta t)$ as discussed later.
    
    The statistical correlation function is the anti-commutator of chromo-electric field operators,
    \begin{align} \label{eq: statistical_correlators}
        \left\langle EE \right\rangle_{jk} (t, t', \mbf x, \mbf x') & = \frac{1}{2} \sum_a \left\langle \left\{ \hat{E}_a^j(x), \hat{E}_a^k(x') \right\}  \right\rangle\,.
    \end{align}
    Here the indices $j, k = 1, 2, z$ correspond to the spatial components, and $x \equiv (t, \textbf{x})$ denotes the spacetime coordinates. Within the classical approximation, the expectation value of an anti-commutator is evaluated by computing the product of classical fields, which simplifies the extraction of the correlation function. 
    In momentum space, such correlators can be evaluated as
    \begin{align} \label{eq: Heisenberg_product}
        \langle EE \rangle_\alpha (t, t', p) &= \frac{1}{V} \sum_a \sum_{j=1}^2\left\langle E_{a, \alpha}^j (t, \textbf{p}) (E_{a, \alpha}^j (t', \textbf{p}))^* \right\rangle, \\
        \langle EE \rangle_z (t, t', p) &= \frac{1}{V} \sum_a \left\langle E_{a}^z (t, \textbf{p}) (E_{a}^z (t', \textbf{p}))^* \right\rangle.
    \end{align}
    Here the index $\alpha = L, T$ denotes gluonic polarizations of the field, which are projections parallel (longitudinal) and transverse to the momentum \textbf{p}. Similarly, the $z$-polarization, i.e., the projection along the $z$-direction, corresponds to the scalar sector. For brevity, we define the sum over the gluonic polarizations
    \begin{align}
        \langle EE \rangle (t, t', p) = \langle EE \rangle_T (t, t', p) + \langle EE \rangle_L (t, t', p)\,,
    \end{align}
    since it is often used in our discussion. It is closely related to the correlator introduced in \eq \eqref{eq: statistical_correlators} for $\mbf x = \mbf x'$ (and similarly for $\langle EE \rangle_z$) that we abbreviate as
    \begin{align}
        \langle EE \rangle (t, t') &\equiv \sum_{j=1} ^2\left\langle EE \right\rangle_{jj} (t, t', \mbf x, \mbf x) \\
        &= \int \frac{\mathrm{d}^2p}{(2\pi)^2} \langle EE \rangle (t, t', p)\,.
    \end{align}

    To compute the statistical correlators in frequency space, the central time $\Bar{t} = (t+t')/2$ and relative time $\Delta t = t' - t$ are introduced. In analogy to previous studies \cite{Boguslavski:2021buh, Boguslavski:2018beu}, we approximate the correlator $\langle EE \rangle (\Bar{t}, \omega)$ as
    \begin{align}
        \langle EE \rangle (\Bar{t}, \omega) & = \int_{-\infty}^\infty \mathrm{d}\Delta t \ e^{-i \omega \Delta t} \langle EE \rangle (\Bar{t} + \Delta t / 2, \Bar{t} - \Delta t / 2) \nonumber \\
        & \!\!\!\!\!\! \approx
        2 \int_\mathrm{0}^{\Delta t_\mathrm{max}} \mathrm{d}\Delta t \ \cos(\omega \Delta t) \langle EE \rangle (t + \Delta t, t),
        \label{eq: FT_approx_BroadExc}
    \end{align}
    where we assumed a separation in time scales $\Delta t_{\mathrm{max}} \ll t \approx \Bar{t}$. This is justified for sufficiently late times $t$ for our systems that we study in a self-similar scaling regime \cite{Boguslavski:2019fsb} and thus we will often write $\langle EE \rangle (t, \omega)$ instead of $\langle EE \rangle (\Bar{t}, \omega)$.
    
    The second type of two-point functions we will discuss are spectral functions and their time derivatives. Contrary to the statistical correlation functions, these are defined as the commutator of two field operators,
    \begin{align}
        \rho_{jk} (t, t', \mbf x, \mbf x') & = i \sum_a \left\langle \left[ \hat{A}^a_j(x), \hat{A}^a_k(x') \right] \right\rangle \nonumber \\
        \dot{\rho}_{jk} (t, t', \mbf x, \mbf x') & = i \sum_a \left\langle \left[ \hat{E}^a_j(x), \hat{A}^a_k(x') \right] \right\rangle \label{eq: rho_definition}
    \end{align}
    and are related to the retarded propagator via $G^R_{jk}(t, t', p) = \theta(t-t') \rho_{jk} (t, t', p)$. They have been computed in 3+1D and 2+1D gauge plasmas far from equilibrium using a gauge-covariant linear response framework for classical-statistical lattice theories in \cite{Kurkela:2016mhu, Boguslavski:2018beu, Boguslavski:2021buh}.
    
    Specifically in \cite{Boguslavski:2021buh}, it was numerically found that the $\langle EE \rangle_\alpha$ correlators of 2+1D systems are related to the (dotted) spectral functions for the same polarization $\alpha = L, T, z$ via a generalized fluctuation-dissipation relation
    \begin{equation} \label{eq: fluct-diss-rel}
        \frac{\langle EE \rangle_\alpha (t, \omega, p)}{\langle EE \rangle_\alpha (t, t, p)} \approx \frac{\dot{\rho}_\alpha (t, \omega, p)}{\dot{\rho}_\alpha (t, t, p)}. 
    \end{equation}
    This relation allows us to connect our results for the form of $\langle EE \rangle$ with the spectral function using the corresponding {\em equal-time} correlations as normalization factors.
    
    Finally, we comment on the gauge (in-)variance of these correlation functions. Due to the employed temporal gauge $A_0=0$, the correlator $\langle EE \rangle (t, t')$, and as a consequence also $\langle EE \rangle (t, \omega)$, can be written in a manifestly gauge-invariant way \cite{Ipp:2020mjc},
    \begin{align}
        \langle EE \rangle_{jk} (t, t') = 2 \, \mathrm{Tr}\,\langle E^j(t) W_0(t, t') E^k(t') W_0(t', t) \rangle,
    \end{align}
    where all quantities are at the same site $\mbf x$. Here, $W_0$ are temporal Wilson lines that reduce to unit matrices in the temporal gauge, and $E^j$ are chromoelectric fields in the fundamental representation. In contrast, (polarized) correlators that depend on momenta like $\langle EE \rangle_\alpha (t, t', p)$ are not gauge invariant. We note that in the temporal gauge, a residual (spatial) gauge freedom remains. We again impose the Coulomb-like gauge $\partial_j A_j = 0 \mid_t$ at the time $t$ before computing $p$-dependent correlation functions.

\subsection{Initial conditions and self-similarity} \label{subsec:selfsim_ic}
    We employ the same Gaussian initial conditions as in \cite{Boguslavski:2021buh, Boguslavski:2019fsb}: The fields are initialized according to the single-particle occupation number distribution function $f(t, \textbf{p})$ at initial time $Qt=0$,
    \begin{equation} \label{eq: init_distrib_func}
        f(t=0, p)=\frac{n_0\,Q}{g^2} e^{-\frac{p^2}{2Q^2}},
    \end{equation}
    where we set the amplitude $n_0 = 0.1$ and use the characteristic initial momentum scale $Q$ as a reference scale for all dimensionful quantities in our study. By rescaling gauge and chromo-electric fields with $g$, the simulations become independent of the coupling. This is equivalent to suppressing vacuum contributions of the full quantum theory for $g \to 0$ while keeping $g^2 f$ fixed.
    
    In \cite{Boguslavski:2019fsb}, it was found that both 2D and Glasma-like models exhibit universal dynamics. The shared gluonic classical attractor is characterized by a self-similar evolution of the system's distribution function $f(t, p) \approx (Q t)^\alpha f_s((Q t)^\beta p)$ with universal scaling exponents $\alpha = 3\beta$ and $\beta = -1/5$. It is independent of the exact initial form and has the physical interpretation of an energy cascade. In particular, the conserved energy density within a hard-momentum region $\sim \int d^2 p\,p f(t,p)$ is transported further to higher momenta in a self-similar way. Such self-similarity is also evident in several of the system's observables. For instance, a typical hard momentum scale where the energy density is dominated increases as $\Lambda (t) \sim Q(Qt)^{-\beta}$. In contrast, the typical soft momenta are characterized by the Debye mass (or plasmon frequency $\wplas \sim m_D$) and decrease with time, $m_D(t) \sim Q (Qt)^{\beta}$. More generally, the time $t$ scaling of different observables indicates which scales dominate in a given process.
    
    We revisit scaling in \se \ref{subsec:k_selfsim} where we briefly discuss the time $t$ scaling properties of the heavy-quark diffusion coefficient $\kappa(t, \Delta t)$. We emphasize that $t$ scaling is specific to the self-similar states we consider while the $\Delta t$ evolution can be more general since it is a consequence of the structure of correlation functions in frequency space. We will use this subtle difference in \se \ref{sec:Glasma} to draw conclusions for $\kappa$ in the Glasma. 


\section{Heavy-quark diffusion in 2+1D systems} \label{sec:hq_diff_in_2d}
    In this section, we introduce the heavy-quark diffusion coefficient in 2+1 dimensional theories. We distinguish gluonic and scalar sectors, $\kappa(t,\Delta t)$ and $\kappa_z(t,\Delta t)$, and introduce different definitions that are equivalent in certain limits to conduct consistency checks of our analysis. We unveil general qualitative features in the $\Delta t$ evolution of $\kappa$ and $\kappa_z$, which will be the focus of the subsequent sections. Finally, we observe a self-similar evolution in time $t$ that originates from the underlying attractor state \cite{Boguslavski:2019fsb} and provide details of the extracted scaling behavior in \app \ref{app:self_sim_parameters}.

\subsection{Heavy-quark diffusion} \label{subsec:kappa_theory}
    The motion of a heavy quark represents a probe for the transport properties of the medium, a characteristic quantity of which is the heavy-quark momentum diffusion coefficient $\kappa$ \cite{Moore:2004tg, Rapp:2018qla}. In this work, it is defined as
    \begin{equation}\label{eq:kappa}
        2 \kappa (t, \Delta t) = \frac{g^2}{N_c} \int_{-\infty}^{\infty} \frac{\mathrm{d}\omega}{2\pi} \frac{\sin(\omega \Delta t)}{\omega} \langle EE \rangle(t, \omega), 
    \end{equation}
    in accordance with \cite{Boguslavski:2020tqz} and is sometimes referred to as $2\kappa_{\mathrm{def}}$ to distinguish it from other definitions that we introduce in the following. This definition, involving an integral over frequency space, allows us to study its relation to the excitation spectrum directly. In the literature, similar definitions for the diffusion coefficient have been used for heavy quarks and quarkonia, where the limit $\Delta t \to \infty$ is usually part of the definition \cite{Brambilla:2016wgg}. 
    
    In the study of the Glasma and over-occupied gluon plasmas, the closely related definitions 
    \begin{align}
        2\kappa^< (t, \Delta t) &= \frac{g^2}{N_c} \int_{t}^{t + \Delta t} \mathrm{d}t' \langle EE \rangle(t + \Delta t, t'), \label{eq: k_lt} \\
        2\kappa^> (t, \Delta t) &= \frac{g^2}{N_c} \int_{t}^{t + \Delta t} \mathrm{d}t' \langle EE \rangle(t, t'),   \label{eq: k_gt}
    \end{align}
    have been used to characterize the momentum diffusion of heavy quarks \cite{Ipp:2020mjc, Boguslavski:2020tqz}. The first expression is directly linked to the momentum broadening of a heavy quark with respect to $\Delta t$ 
    \begin{align}
        \frac{\mathrm{d}}{\mathrm{d}\Delta t} \langle p^2(t, \Delta t) \rangle = 2\kappa^<(t,\Delta t).
        \label{eq:psqr_relate}
    \end{align}
    For sufficiently small $\Delta t$, \eq\eqref{eq: k_lt} can be approximated by dropping the $\Delta t$-dependence of the integrand. This was employed in \cite{Boguslavski:2020tqz} and leads to $2\kappa^>$. Using the approximation \eqref{eq: FT_approx_BroadExc} for the Fourier transform $\langle EE \rangle (t, \omega)$ in the definition \eqref{eq:kappa}, one finds that $\kappa$ agrees with $\kappa^>$. Hence, we obtain
    \begin{align}
        \label{eq:kappa_agree}
        \kappa(t, \Delta t) \approx \kappa^>(t, \Delta t) \approx \kappa^< (t, \Delta t).
    \end{align}
    We will compare the expressions for $\kappa$ numerically in \se \ref{subsec:k_computation}.

    More precisely, the expressions in \eqs \eqref{eq:kappa} - \eqref{eq: k_gt} correspond rather to the quarkonium transport coefficient, which is in general different from the heavy quark momentum diffusion coefficient \cite{Scheihing-Hitschfeld:2022xqx}. However, it has recently been shown in classical-statistical simulations of the Glasma \cite{Avramescu:2023qvv} and of 3+1 dimensional gluonic plasmas \cite{Pandey:2023dzz} that these definitions coincide with the momentum diffusion of heavy quarks. 

    We emphasize that we consider the heavy-quark diffusion coefficient in non-equilibrium systems, which therefore depends on both times $t$ and $\Delta t$. In thermal equilibrium, there is no dependence on $t$ due to time-translation invariance and one typically defines the heavy-quark diffusion coefficient in the limit \cite{Caron-Huot:2007rwy, Laine:2009dd}
    \begin{align} \label{eq: Kappa_therm}
        \kappa_\infty^\mathrm{therm} = \lim_{\Delta t \to \infty} \kappa^\mathrm{therm} (\Delta t)\,.
    \end{align}
    In analogy, we extract for sufficiently large $\Delta t$
    \begin{align} 
        \label{eq:kappa_infty}
        \kappa_\infty (t) = \lim\limits_{1/Q \ll \Delta t \ll t} \kappa (t, \Delta t),
    \end{align}
    which is similar to the definition in \cite{Boguslavski:2020tqz} for non-thermal systems and can exhibit a residual $t$-dependence. 
   
    We can also deduce the behavior at small $Q \Delta t \lesssim 1$. In this case, the correlation function of chromo-electric fields barely changes $\langle EE \rangle \approx \langle EE \rangle(t, \Delta t \to 0)$, which leads to linear growth of $\kappa$
    \begin{align}
        \label{eq:kappa_early}
        2\kappa(t, \Delta t) \approx \frac{g^2}{N_c}\, \langle EE \rangle(t,0)\;\Delta t .
    \end{align}
    Using the relation \eqref{eq:psqr_relate}, one retrieves the characteristic behavior at early times $\langle p^2(t, \Delta t) \rangle \propto \Delta t^2$, which is also visible in 3+1 dimensional plasmas \cite{Boguslavski:2020tqz} and the Glasma \cite{Khowal:2021zoo, Avramescu:2023qvv}.
    
    In the discussion above, we have focused on the heavy-quark diffusion coefficient $\kappa$ of the gluonic contribution. All arguments are valid for the scalar sector as well where we substitute with the analogous quantities $\langle EE \rangle \mapsto \langle EE \rangle_z$, $2 \kappa \mapsto \kappa_z$, and $\langle p^2 \rangle \mapsto \langle p_z^2 \rangle$.

\subsection{Perturbative estimate of \texorpdfstring{$\boldsymbol{\kappa_\infty}$}{k}} \label{subsec:kappa_infty_pert}
    Besides the computation of $\kappa$, we are also interested in an estimate of how $\kappa_\infty(t)$ could evolve in time using solely HTL perturbation theory (see \app \ref{app:HTL} for details). Although, strictly speaking, it is not applicable due to the existence of important non-perturbative features in the correlation functions, it is still a useful tool that can sometimes capture qualitative properties and that allows us to compare our numerical results. 

    Our starting point is \eq \eqref{eq:kappa}. In the limit $\Delta t \to \infty$, one has $\sin(\omega \Delta t)/\omega \to \pi \delta(\omega)$ and the expression simplifies to
    \begin{align}
        2 \kappa_\infty (t) &= \frac{g^2}{2 N_c} \langle EE \rangle(t, \omega{=}0) \nonumber \\
        &= \frac{g^2}{2 N_c} \int \frac{\mathrm{d}^2p}{(2\pi)^2} \langle EE \rangle(t, \omega{=}0, p) \nonumber \\
        &\overset{\mathrm{pert}}{=} \frac{g^2}{2 N_c} \int \frac{\mathrm{d}^2p}{(2\pi)^2} \frac{\langle EE \rangle_L(t, t, p)}{\dot \rho_L(t, t, p)}\, \dot \rho_L(t, \omega{=}0, p).
    \end{align}
    In the last line, we have used the generalized fluctuation-dissipation relation \eqref{eq: fluct-diss-rel} and the expectation from HTL perturbation theory that only the longitudinal Landau cut contribution $\dot \rho_L(t, \omega{=}0, p) = 2p\, m_D^2/(m_D^2 + p^2)^2$ does not vanish at $\omega = 0$. One can also use the sum rule in \eq \eqref{eq: L_sumrule}, i.e., $\dot \rho_L(t, t, p) = m_D^2/(m_D^2+p^2)$. Moreover, it has been found in \re \cite{Boguslavski:2019fsb} that the equal-time correlator follows approximately 
    \begin{align}
        g^2\langle EE \rangle_L(t, t, p) \approx g^2 T_*(t)\,\frac{d_A\,m_D^2(t)}{m_D^2(t) + p^2}
        \label{eq:eff_T}
    \end{align}
    for $p \lesssim Q$, where the transverse correlator becomes constant (cf., upper panel of the later \fig \ref{fig: equal-time_corrs}) and defines the effective temperature as 
    \begin{align}
        g^2 T_*(t) = g^2\langle EE \rangle_T(t, t, p)/d_A\,.
    \end{align}
    With these, we obtain
    \begin{align}
        2 \kappa_\infty^{\mathrm{pert}} (t) 
        &= \frac{g^2 T_*(t)\,d_A}{2 N_c} \int^{Q} \frac{\mathrm{d}^2p}{(2\pi)^2} \, \frac{2 p\, m_D^2(t)}{(m_D^2(t) + p^2)^2} \nonumber \\
        &= \frac{N_c^2-1}{8 N_c}\, g^2 T_*(t)\,m_D(t)\;\left(1 + \mathcal O \left(\frac{m_D}{Q}\right)\right).
        \label{eq:kappa_pert}
    \end{align}
    Interestingly, $2 \kappa_\infty^{\mathrm{pert}} (t)$ is dominated by soft modes $p \sim m_D$ and does not include an additional logarithmic dependence on hard and soft momenta as in 3+1D \cite{Moore:2004tg, Caron-Huot:2007rwy, Caron-Huot:2008dyw}. 

    In the considered self-similar regime (see \se \ref{subsec:selfsim_ic}), we have $m_D \sim Q (Qt)^\beta$ and $g^2 T_* \sim Q^2 (Qt)^{2\beta}$. The latter results from the scaling properties of $f(t,p)$ and $f \sim \langle EE \rangle_T(t, t, p)/p$. With $\beta = -1/5$, we obtain the perturbative scaling estimate
    \begin{align}
        2 \kappa_\infty^{\mathrm{pert}} (t) \sim g^2 T_*(t)\,m_D(t) \sim Q^3 (Qt)^{-3/5}.
    \end{align}
    
    For the scalar sector, the correlators have no contribution at $\omega=0$ in the HTL framework as $\langle EE \rangle_z(t, \omega{=}0, p) = 0$, and thus, we have 
    \begin{align}
        \label{eq:kappa_z_pert}
        \kappa_{z, \infty}^{\mathrm{pert}} = 0.
    \end{align}

\subsection{Numerical computation of \texorpdfstring{$\boldsymbol{\kappa (t, \Delta t)}$}{k}} \label{subsec:k_computation}
    We proceed with a discussion of our numerical results for the heavy-quark momentum diffusion coefficients $2\kappa$ and $\kappa_z$ as introduced in \se \ref{subsec:kappa_theory}. They are obtained for the 2D theory on an $N_s^2 = 1024^2$ lattice with lattice spacing $Qa_s = 1/8$ and in the Glasma-like theory on a $512^2$ lattice with spacing $Qa_s = 1/4$. The time step is $a_t = 0.05\ a_s$ and we average over $420-1120$ independent simulations. If not stated otherwise, we compute our observables at $Qt =500$.
   
    \begin{figure}[t!]
        \centering
            \includegraphics[width=0.4\textwidth]{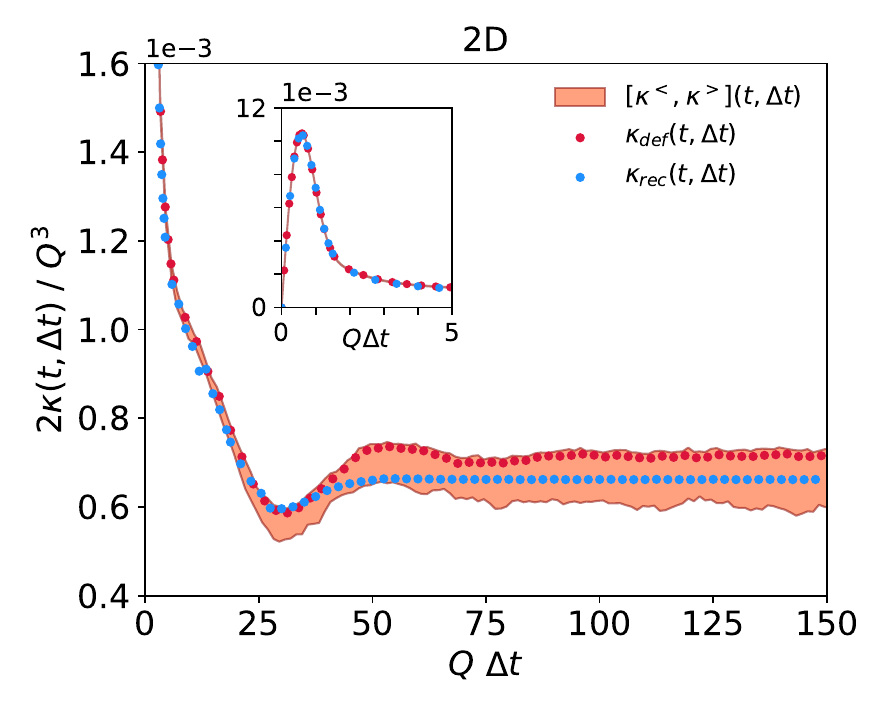}
            \includegraphics[width=0.4\textwidth]{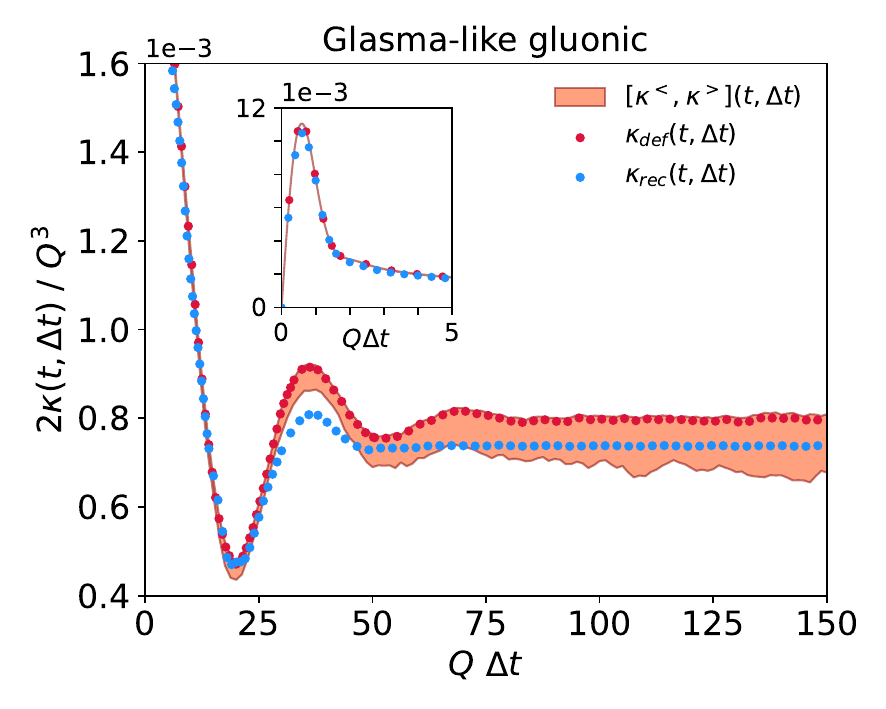}
            \includegraphics[width=0.4\textwidth]{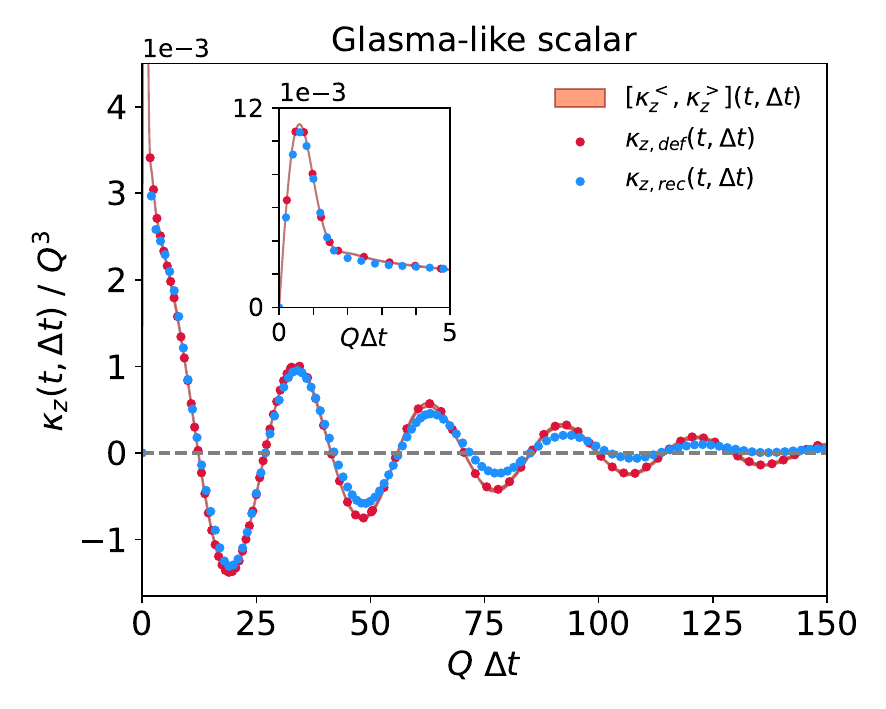}
            \caption{Heavy-quark diffusion coefficient $2\kappa$ for the 2D system ({\em top}), the gluonic sector of the Glasma-like system ({\em center}) and its scalar contribution $\kappa_z$ ({\em bottom}) as functions of $\Delta t$ at $Qt=500$. The insets depict early times that have been partially cut in the main panels. We also compare different expressions $\kappa^{\lessgtr}$ and $\kappa_{\mrm{def}}$ given by \eqs \eqref{eq:kappa}-\eqref{eq: k_gt} and the reconstructed curves $\kappa_{\mrm{rec}}$ that will be discussed in \se \ref{sec:spectral_reconst}.
            }
        \label{fig:kappa_vs_bounds}
    \end{figure}
    \begin{figure}[t!]
        \centering
            \includegraphics[width=0.4\textwidth]{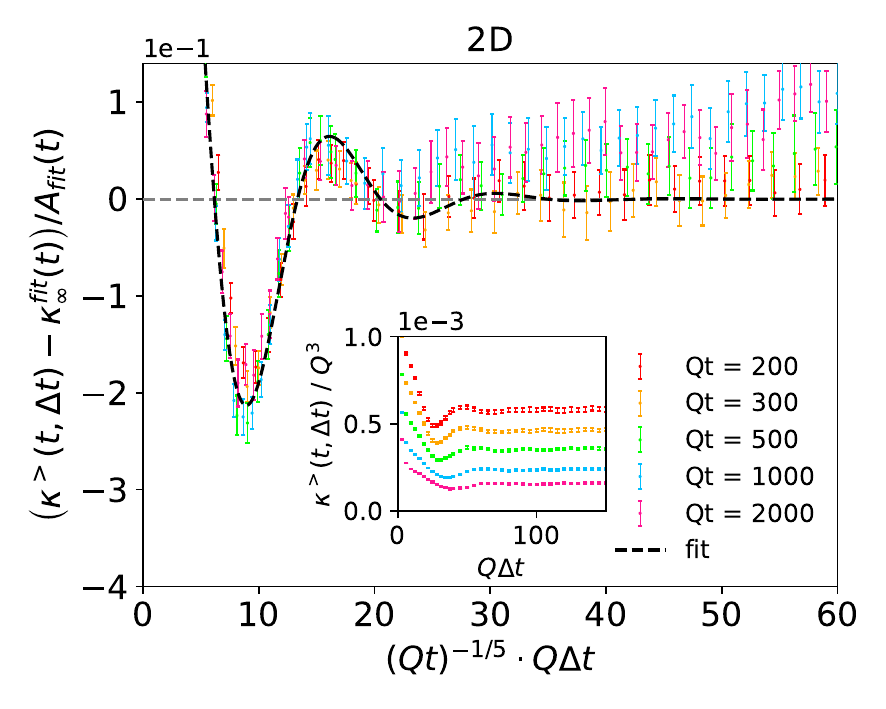}
            \includegraphics[width=0.4\textwidth]{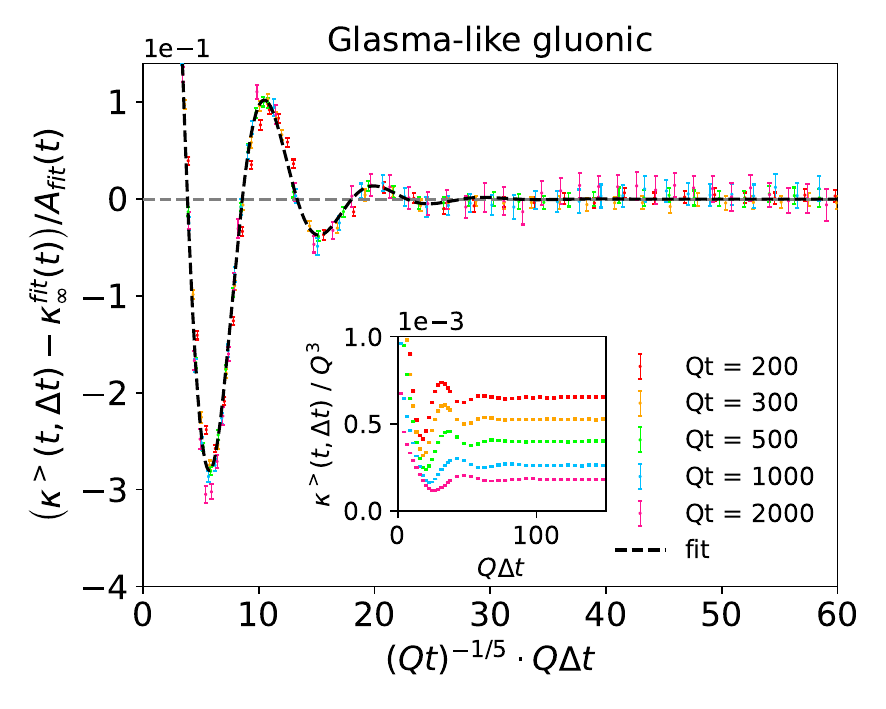}
            \includegraphics[width=0.4\textwidth]{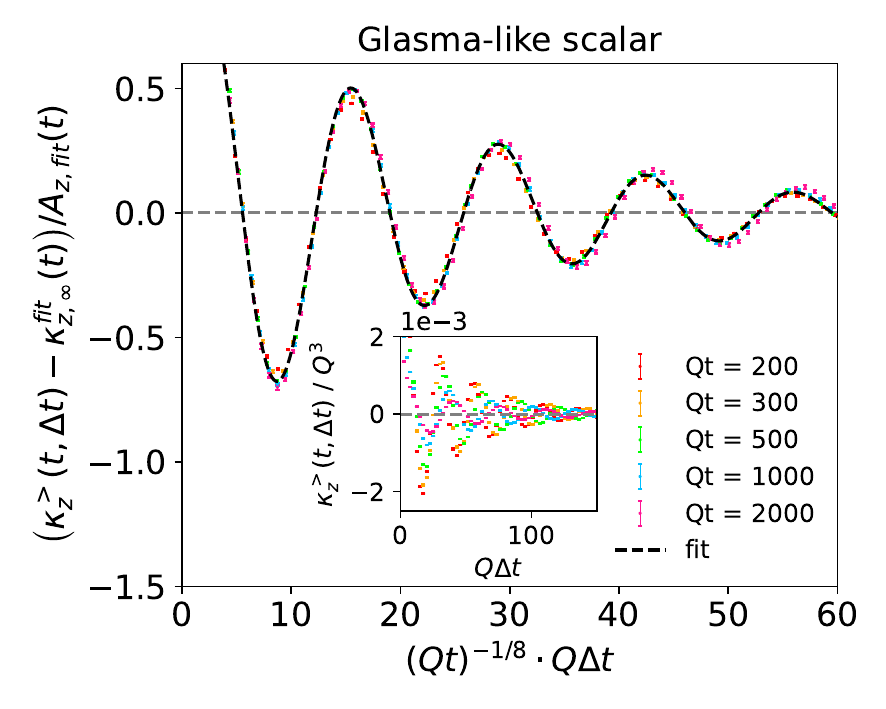}
            \caption{
                Overview of the scaling behavior of  $\kappa$ in the 2D theory ({\em top}) and $\kappa$ ({\em center}) and $\kappa_{z}$ ({\em bottom}) in the Glasma-like system for various times $Qt$. The main panels show rescaled curves using \eq \eqref{eq:rescaling} as functions of rescaled $\Delta t$ with factor $Q(Qt)^{-\sigma_{\mathrm{gl}}} \sim \omega_\mathrm{pl}$ for the gluonic and $Q(Qt)^{-\sigma_z}$ for the scalar contributions. The black-dashed line shows the rescaled fit in \eq\eqref{eq: damp_HO} at time $Qt=500$. The original curves are shown in the insets. 
            }
        \label{fig:kappa_OV}
    \end{figure}

    The upper two panels of \fig \ref{fig:kappa_vs_bounds} show the gluonic coefficient $2\kappa$ as a function of $Q \Delta t$ for the 2D and Glasma-like theories. The bottom panel shows $\kappa_z$ for the latter. The definitions \eqref{eq:kappa}, \eqref{eq: k_lt} and \eqref{eq: k_gt} are denoted as $\kappa_{\mathrm{def}}$, $\kappa^<$ and $\kappa^>$, respectively. In each panel, one observes that these quantities coincide at early times (insets) and evolve in a similar way to each other at later times.%
    \footnote{
        One observes that $\kappa^>$ coincides with $\kappa_\mathrm{def}$, which is a consequence of the approximation in \eq \eqref{eq: FT_approx_BroadExc} to extract $\langle EE \rangle (t, \omega)$. 
    }
    Their comparison allows an estimation of the systematic uncertainties and is highlighted by the orange band between $\kappa^<$ and $\kappa^>$. The uncertainty is generally small in comparison to their magnitude but is seen to grow with time for $2\kappa$.

    The gluonic coefficient $2\kappa$ exhibits damped oscillations in $\Delta t$ after a short but steep rise as shown in the insets. In particular, one retrieves the predicted linear behavior of \eqref{eq:kappa_early} for $Q \Delta t \lesssim 1$. The subsequent evolution is similar for the gluonic sectors of the theories but is qualitatively different for the scalar sector. In particular, $2 \kappa$ is always positive, strongly damped, and approaches a constant $\kappa_\infty$ for large $\Delta t$ as expected from \eq \eqref{eq:kappa_infty}. The latter implies heavy-quark diffusion with a constant momentum growth rate \eqref{eq:psqr_relate}, which can be approximated by a simple Langevin equation for its momentum at sufficiently late times. In contrast, $\kappa_z$ exhibits weakly damped oscillations around zero, i.e., involves positive as well as negative values. Such non-diffusive behavior implies momentum broadening and narrowing and is the main qualitative difference to the gluonic sector. The beginning of such an evolution has been observed in the Glasma \cite{Avramescu:2023qvv} and will be further discussed in \se \ref{sec:Glasma}. 

    To understand the similarities and differences of $\kappa$ and $\kappa_z$, we reconstruct their evolution using correlation functions and show the reconstructed curves in \fig \ref{fig:kappa_vs_bounds} as $\kappa_\mathrm{rec}$. While we elaborate on our reconstruction approach and its impact in \ses \ref{sec:spectral_reconst} and \ref{sec:properties_reconst}, we note here that the curves reproduce the evolution in each panel with good accuracy. This allows us to obtain a detailed understanding of the observed features in the $\Delta t$ evolution.

\subsection{Self-similar \texorpdfstring{$\boldsymbol{t}$}{t}-scaling of \texorpdfstring{$\boldsymbol{\kappa (t, \Delta t)}$}{k}} \label{subsec:k_selfsim}

    We now turn to the overall $t$ dependence of $\kappa$ and $\kappa_z$. Similarly to the 3+1 dimensional case \cite{Boguslavski:2020tqz}, their evolution depends primarily on the dynamics of the distribution functions of the gluonic and scalar sectors. As we discussed in \se \ref{subsec:selfsim_ic}, the systems we consider follow a self-similar evolution with universal scaling exponents. Hence, self-similarity in $t$ can also be expected for the diffusion coefficients as functions of $\Delta t$. To see this, we fit $\kappa$ and $\kappa_{z}$ at fixed $t$ to a damped harmonic oscillator (in $\Delta t$) according to
    \begin{align} \label{eq: damp_HO}
        &\kappa_\mathrm{fit}(t, \Delta t) \\
        &~~~ \approx \kappa_\infty^\mathrm{fit} (t) + A_\mathrm{fit}(t) \cos(\omega_\mathrm{fit}(t) \Delta t - \phi_\mathrm{fit})\, e^{- \gamma_\mathrm{fit}(t) \Delta t}. \nonumber
    \end{align}
    The offset $\kappa_\infty^\mathrm{fit}$, amplitude $A_\mathrm{fit}$, frequency $\omega_{\mathrm{fit}}$ and damping $\gamma_\mathrm{fit}$ are introduced as $t$ dependent parameters for a range of times $Qt = 200, 300, 500, 1000,$ and $2000$. In the fitting procedure, the phase shift $\phi_\mathrm{fit}$ is fixed to a time-independent constant for each theory and sector (see \app \ref{app:self_sim_parameters} for the values).

    To analyze the universality of the evolution in \fig \ref{fig:kappa_OV}, we subtract the late-time diffusion coefficient, normalize the quantity according to
    \begin{equation} \label{eq:rescaling}
        \frac{\kappa (t, \Delta t) - \kappa_{\infty}^\mrm{fit}(t)}{A_\mrm{fit}(t)},
    \end{equation}
    and show the curves as functions of the rescaled time $(Qt)^{-\sigma} \cdot Q \Delta t$. The insets show the original transport coefficients. One observes that, while the original curves are rather distinct, once rescaled, they coincide in every panel within statistical uncertainties. This very good agreement demonstrates self-similarity and a universal functional form for $\kappa$ and $\kappa_z$ as functions of $\Delta t$. 

    Let us now discuss and interpret the rescaling of the $\Delta t$ time axis in \fig \ref{fig:kappa_OV}. For the gluonic diffusion coefficients, we find that the exponent $\sigma_{\mathrm{gl}} = 1/5$ leads to a self-similar evolution for both, the  2D and Glasma-like theories. This value suggests a tight connection to the time evolution of the plasmon frequency that also follows $\omega_\mathrm{pl} \sim Q(Qt)^{-1/5}$ \cite{Boguslavski:2019fsb, Boguslavski:2021buh}. For a 3+1D theory, it was indeed shown that $\omega_{\mathrm{fit}}$ and $\omega_\mathrm{pl}$ are even quantitatively close \cite{Boguslavski:2020tqz}. Similar to that study, our interpretation of the systems considered here is that the late-time evolution and the emerging oscillations originate from properties of low-momentum excitations $p \lesssim \wplas$. Interestingly, we find $\sigma_z \approx 1/8$ for $\kappa_z$ of the Glasma-like theory. Hence, it exhibits a different scaling behavior although it has been shown that the dispersion relation of the $z$ excitation scales with $\wplas$ at low momenta \cite{Boguslavski:2021buh}. We have not yet found a coherent explanation for $\sigma_z$. 

    \begin{figure}[t!]
        \centering
        \includegraphics[width=0.49\textwidth]{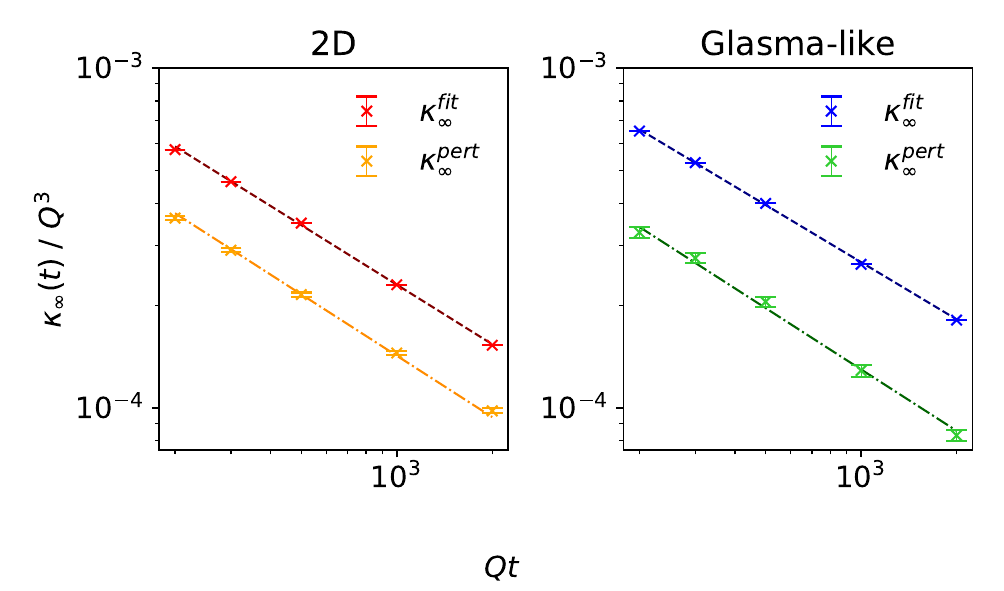}
        \caption{
            The fitted $\kappa_\infty^\mathrm{fit}$ from \eq \eqref{eq: damp_HO} as a function of time $t$ for the gluonic sector of the 2D (\emph{left}) and the Glasma-like (\emph{right}) theory. The scalar coefficient for the latter does not exhibit an offset and is therefore not shown. The dashed curves show power laws $\sim Q^3 (Qt)^{-\sigma_\kappa}$ with the fitted exponents \eqref{eq:kappa_fit_exp}. We also show the perturbatively expected $\kappa_\infty^\mathrm{pert}(t)$ from \eq\eqref{eq:kappa_pert}. It follows a similar power law with scaling exponent $3/5$ but significantly undershoots our non-perturbative results. 
        }
        \label{fig:kappa_inf}
    \end{figure}

    We proceed by studying the evolution of the fit parameters $A_\mrm{fit}(t)$, $\omega_\mrm{fit}(t)$ and $\gamma_\mrm{fit}(t)$, before focusing on $\kappa_\infty^{\mathrm{fit}}(t)$. While a detailed discussion of our results is given in \app \ref{app:self_sim_parameters}, we summarize our main observations. All fit parameters are well described by power laws $\sim Q(Qt)^{-\sigma}$. The scaling exponents $\sigma_\omega$ and $\sigma_\gamma$ of $\omega_\mrm{fit}(t)$ and $\gamma_\mrm{fit}(t)$ roughly agree with the $\Delta t$-scaling exponents $1/5$ and $1/8$ employed in \fig \ref{fig:kappa_OV} for $\kappa(t,\Delta t)$ and $\kappa_{z}(t,\Delta t)$, respectively. For the gluonic quantity $\kappa$, we further observe the relation $\omega_\mrm{fit}\approx\omega_\mrm{pl}\sim Q(Qt)^{-1/5}$. We find that the gluonic decay rate scales similarly as $\gamma_\mrm{fit} \sim \omega_\mrm{pl}$, suggesting important contributions to $\kappa$ from the low momentum region. In contrast to $\sigma_\omega$, a simple interpretation of the amplitude scaling coefficient $\sigma_A\approx 0.6-0.7$ is more difficult to achieve because the approximate scaling behavior may originate from an interplay of different momentum scales. Therefore, a description of $\kappa$ based on the excitation spectrum is more promising and is the focus of the subsequent sections.

    Finally, we discuss our results for the late-$\Delta t$ heavy-quark diffusion coefficient $\kappa_\infty$ of \eq \eqref{eq:kappa_infty} that emerges as a plateau, which we compare to the purely perturbative expectations of \se \ref{subsec:kappa_infty_pert}. The scalar diffusion coefficient vanishes, $\kappa_{z, \infty} = 0$, since $\kappa_z$ oscillates around approximately zero, which is also the case for the perturbative value. The gluonic sectors of both models are shown in \fig \ref{fig:kappa_inf}, where we depict the fitted values $\kappa_\infty^{\mathrm{fit}}$ for different times $t$. We observe that $\kappa_\infty^{\mathrm{fit}}(t)$ decreases with time in a similar way in both theories. To quantify its time dependence, we employ power law fits $\kappa_\infty^{\mathrm{fit}}\sim Q^3 (Q t)^{-\sigma_\kappa}$ yielding
    \begin{align}
        \sigma^\mathrm{2D}_\kappa = 0.58 \pm 0.01\,, \qquad \sigma^\mathrm{2D+sc}_\kappa = 0.56 \pm 0.01\,.
        \label{eq:kappa_fit_exp}
    \end{align}
    These values are close to the scaling exponent $3/5$ for $\kappa_\infty^{\mathrm{pert}} (t)$ of \se \ref{subsec:kappa_infty_pert} that is expected if solely HTL perturbation theory was applied. For a more quantitative comparison, we evaluate the perturbative result \eqref{eq:kappa_pert} for which we compute the effective temperature $g^2 T_*$ and Debye mass $m_D$ using \eq \eqref{eq:eff_T} and the self-consistent mass formula \eqref{eq: asympt_mass}. Our results for $\kappa_\infty^{\mathrm{pert}} (t)$ are added to \fig \ref{fig:kappa_inf} together with power laws $\kappa_\infty^{\mathrm{pert}} (t) \sim Q^3 (Qt)^{-3/5}$. We observe that, while the self-similar scaling matches with good accuracy, the perturbative values are considerably smaller by a factor of 2 to 3 than our simulation results $\kappa_\infty^{\mathrm{fit}}(t)$. This indicates that $\kappa_\infty^{\mathrm{pert}}(t)$ misses crucial contributions that provide at least 50\% of the actual value.

    Indeed, $\kappa_\infty^{\mathrm{pert}} (t)$ takes only the Landau damping region into account and assumes quasi-particles with an infinite lifetime that do not contribute to its value. Thus, the perturbative calculation misses any features of the excitations beyond these assumptions. The mismatch with our results, therefore, indicates that HTL perturbation theory cannot correctly describe 2+1 dimensional non-Abelian plasmas and an accurate description requires the inclusion of non-perturbative features. That HTL perturbation theory breaks down in such systems is not surprising since soft contributions become as important as hard ones in loop diagrams \cite{Boguslavski:2019fsb, Boguslavski:2021buh}. One of the main results of our work is that some of the required non-perturbative contributions stem from a new transport peak and broad gluonic excitations that influence heavy-quark diffusion substantially, as we will show in the following sections.


\section{Reconstructing \texorpdfstring{$\boldsymbol{\kappa (t, \Delta t)}$}{k} from correlation functions} \label{sec:spectral_reconst}
    In this section, we discuss the spectral and statistical correlation functions of the considered systems and show that these gauge-fixed unequal-time correlation functions can reconstruct the $\Delta t$ evolution of the gauge-invariant heavy-quark diffusion coefficients \eqref{eq:kappa}. In a similar spirit as for 3+1D in \cite{Boguslavski:2020tqz}, this will enable us to assess the physical impact of different features of the underlying excitation spectrum in \se\ref{sec:properties_reconst}.

\subsection{Properties of the spectral and statistical functions} \label{subsec:prop_sectral_func}
    Before discussing the reconstruction of $\kappa(t, \Delta t)$, we introduce the relevant unequal-time correlators for this procedure. The spectral function $\rho$ as well as its time derivative $\dot{\rho}$ and statistical correlator $\langle EE \rangle$ were extracted in \cite{Boguslavski:2021buh} using the linear response framework developed in \cite{Kurkela:2016mhu, Boguslavski:2018beu}. We have discussed the definitions and extraction methods in \se \ref{subsec:corr_funcs} and briefly revisit the most important findings in this section.

    In momentum space, we distinguish between the transverse, the longitudinal, and in the case of the Glasma-like theory, the $z$-component of the correlation functions. While in 3+1D, separate polarizations are well described by HTL perturbation theory \cite{Boguslavski:2018beu}, this description partially breaks down in lower dimensional theories because soft modes become important and no scale separation can be assumed \cite{Boguslavski:2019fsb}. This leads to non-perturbative features like broad gluonic excitation peaks of Gaussian form 
    \begin{equation} \label{Gaussfit}
        h_\mathrm{Gauss} = \frac{A}{\gamma}\sqrt{\frac{\pi}{2}} \exp{ \left[ -\frac{1}{2} \left( \frac{\omega - \omega_R}{\gamma} \right)^2 \right]} + \left[ \omega_R \mapsto -\omega_R \right],
    \end{equation}
    where $\left[ \omega_R {\mapsto} -\omega_R \right]$ denotes the previous term with the indicated substitution. However, at sufficiently large momenta some properties, like the (longitudinal) Landau damping of the spectral function, are well-approximated by perturbation theory.

    \begin{figure}[t]
        \centering
        \includegraphics[width=0.49\textwidth]{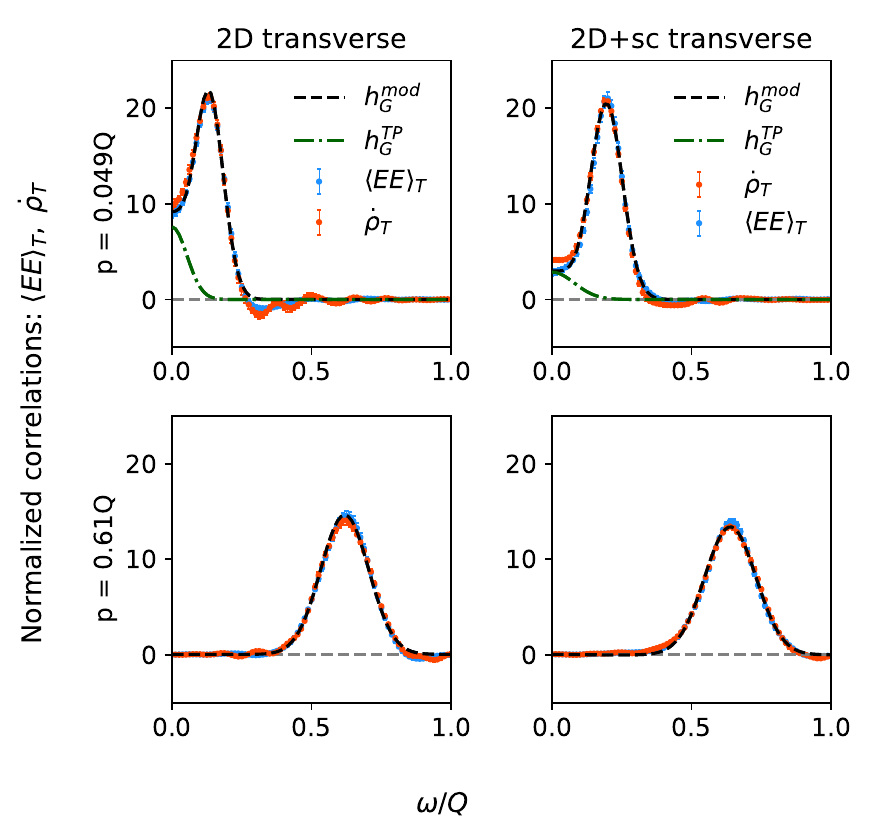}
        \caption{
            The transverse normalized correlation functions $\langle EE \rangle_T (t, \omega) \big/ \langle EE \rangle_T (t, \Delta t{=}0)$ and $\dot{\rho}_T (t, \omega)\big/\dot{\rho}_T (t, \Delta t{=}0)$ are shown for the gluonic sectors of the 2D and Glasma-like theory at representative low ({\emph{top}}) and high ({\emph{bottom}}) momenta. The green dash-dotted curve corresponds to the novel transport peak $h_G^{\mrm{TP}}$ at low momenta and has been incorporated into the modified Gaussian fit to the data $h_G^{\mrm{mod}}$ in \eq \eqref{eq:Gaussfit_mod} (black dashed line).
        }
        \label{fig: trans_comp_SF_Corrs}
    \end{figure}

    \begin{figure}[t]
        \centering
        \includegraphics[width=0.49\textwidth]{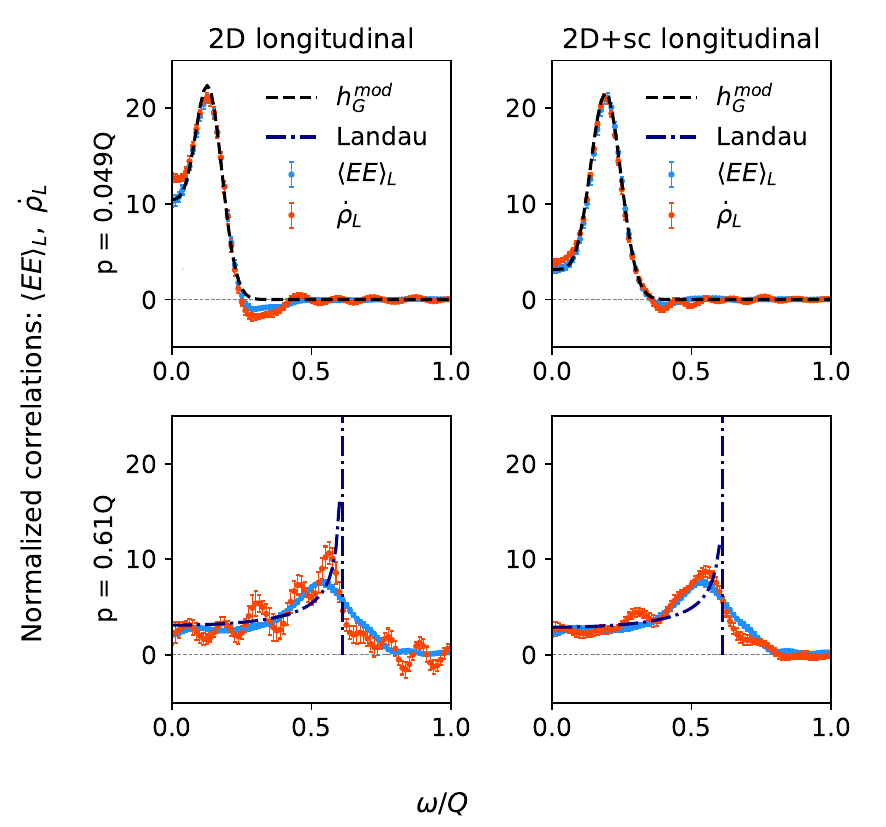}
        \caption{
            The normalized longitudinal correlation functions $\langle EE \rangle_L (t, \omega) \big/ \langle EE \rangle_L (t, \Delta t{=}0)$ and dotted spectral functions $\dot{\rho}_L (t, \omega) \big/ \dot{\rho}_L (t, \Delta t{=}0)$ for the 2D and Glasma-like theories for a small ({\emph{top}}) and high ({\emph{bottom}}) momentum. A Gaussian function (black dashed line) describes the correlator at the small momentum $p \lesssim m_D$, while for $p \gtrsim m_D$, the analytical Landau cut (blue dash-dotted curve from \eq \eqref{eq: L_Landau}) becomes dominant.
        }
        \label{fig: long_comp_SF_Corrs}
    \end{figure}

    We show the normalized transverse and longitudinal spectral functions in \figs \ref{fig: trans_comp_SF_Corrs} and \ref{fig: long_comp_SF_Corrs}, respectively, as functions of frequency $\omega$ at $Qt = 500$. The upper panels correspond to a small representative momentum $p=0.049Q < m_D$ while the lower panels show a hard representative mode $p=0.61Q > m_D$.%
    \footnote{
        The Debye mass $m_D = m_\mathrm{HTL}$ is calculated using the perturbative expression \eqref{eq: asympt_mass} and yields $m_\mathrm{HTL}^\mathrm{2D}=0.14 Q$ and $m_\mathrm{HTL}^\mathrm{2D+sc}=0.21 Q$ for the respective theory at $Q t = 500$.
    }
    One observes that the normalized spectral functions $\dot \rho_{\alpha}$   are in very good agreement with the normalized electric field correlators $\langle EE \rangle_{\alpha}$, confirming the validity of the generalized fluctuation-dissipation relation in \eq \eqref{eq: fluct-diss-rel}. At high momenta, the spectral function can be approximated by HTL perturbation theory (see \app \ref{app:HTL}),
    \begin{equation}
         \label{eq: SF_fit}
        \frac{\dot{\rho}_\alpha (t, \omega, p)}{\dot{\rho}_\alpha (t, t, p)} \approx
        h (\omega, p) + \frac{\omega \ \rho_\alpha^\mathrm{Landau}(t, \omega, p)}{\dot{\rho}_\alpha (t, t, p)}
    \end{equation}
    where $h$ denotes the quasi-particle contribution to the excitations, $\rho_\alpha^\mathrm{Landau}$ is known as the Landau cut defined in \eqs\eqref{eq: T_Landau}-\eqref{eq: phi_Landau}, and the normalization factors $\dot{\rho}_\alpha (t, t, p)$ are determined analytically via the sum rules in \eqs\eqref{eq: T_sumrule} - \eqref{eq: phi_sumrule} and are only different from unity for the longitudinal polarization. Indeed, the Landau cut contribution (dark blue dash-doted line) dominates the longitudinal correlator at $p \gtrsim m_D$ as is visible in \fig \ref{fig: long_comp_SF_Corrs}. For the transverse polarization in \fig \ref{fig: trans_comp_SF_Corrs}, this contribution is suppressed and a Gaussian form $h_\mathrm{Gauss}$ has been shown to accurately capture the excitation peak (black dashed). 
    
    Note that in contrast to this observation, leading order HTL leads to narrow, delta-like quasi-particle peaks or excitations with Lorentzian or Breit-Wigner forms for finite lifetimes. Evidently, the gluonic excitations in the considered 2+1 dimensional plasmas exhibit instead broad excitation peaks of a Gaussian form, particularly at low momenta. This suggests that HTL perturbation theory is not applicable to describe these excitations. 

    Indeed, we highlight the non-vanishing part of $\dot \rho_\alpha$ at vanishing $\omega$ in the figures although the (transverse) spectral function $\rho_T$ as well as $\dot \rho_T \approx \omega \rho_T$ should vanish perturbatively. As noted in \cite{Boguslavski:2021buh}, a Gaussian $h_\mrm{Gauss}$ with finite dispersion relation $\omega_R(p) > 0$ leads to an offset at $\omega = 0$ for broad excitations. However, we find here that an accurate description of the offset requires the existence of an additional transport peak (green dash-dotted curves in \fig \ref{fig: trans_comp_SF_Corrs}) at zero frequency $h_\mrm{G}^\mrm{TP} = h_\mrm{Gauss}(\omega_R{=}0)$ that exists only at low momenta $p \lesssim m_D$. The spectral function $\dot \rho_\alpha$ for both polarizations is thus well described by a modified Gaussian excitation (black dashed curves) 
    \begin{align}
        \label{eq:Gaussfit_mod}
        h_\mrm{G}^\mrm{mod} = h_\mrm{G}^\mrm{TP} + h_\mrm{Gauss}\,.
    \end{align}
    Remarkably, at low frequencies, the offset implies that the original spectral function $\rho_T (t, \omega \to 0, p)$ behaves as $\sim 1/\omega$. In the following sections, we show that this offset has physical implications, which we demonstrate by the explicit reconstruction of the heavy-quark diffusion coefficient.

    In contrast to the gluonic excitations, the scalar excitation peaks at low momenta $p \lesssim m_D$ are narrow, similar to the gluonic peaks observed in 3+1D  \cite{Boguslavski:2021buh} and do not exhibit a finite offset at vanishing momenta. In the same work, it was observed that they coincide with their transverse counterparts for larger momenta, which we will use in our reconstruction strategy. 

\subsection{Reconstruction strategy of \texorpdfstring{$\boldsymbol{\kappa (t, \Delta t)}$}{k}} \label{subsec:numer_reconst_strategy}
    Using the unequal-time correlators of the previous subsection, we can now reconstruct $\kappa(t, \Delta t)$ and interpret its dynamics in terms of properties of the underlying excitation spectrum. The definition of $\kappa$ in \eq\eqref{eq:kappa} allows to rewrite the expressions
    \begin{align}
        &2 \kappa(t, \Delta t)
        = \frac{g^2}{N_c} \int_{\mbf p}\int_{-\infty}^{\infty}
                \frac{\mathrm{d}\omega}{2\pi} \frac{\sin(\omega \Delta t)}{\omega} \sum_{\alpha} \langle EE           \rangle_\alpha (t, \omega, p) \label{eq: SR_EE} \\
        & \approx \frac{g^2}{N_c} \int_{\mbf p} \int_{-\infty}^{\infty}
            \frac{\mathrm{d}\omega}{2\pi} \frac{\sin(\omega \Delta t)}{\omega} 
            \sum_{\alpha} \langle EE \rangle_\alpha (t, t, p) \frac{\dot{\rho}_\alpha (t, \omega, p)}{\dot{\rho}_\alpha (t, t, p)},\label{eq: SR_rho}
    \end{align}
    \begin{align}
        &\kappa_z(t, \Delta t)
            = \frac{g^2}{N_c} \int_{\mbf p} \int_{-\infty}^{\infty}
            \frac{\mathrm{d}\omega}{2\pi} \frac{\sin(\omega \Delta t)}{\omega} \langle EE \rangle_z (t, \omega, p) 
            \label{eq: SR_phi} \\
        &\approx \frac{g^2}{N_c} \int_{\mbf p} \int_{-\infty}^{\infty} \frac{\mathrm{d}\omega}{2\pi} \frac{\sin(\omega \Delta t)}{\omega} \langle EE \rangle_z (t, t, p) \frac{\dot{\rho}_z (t, \omega, p)}{\dot{\rho}_z (t, t, p)}, \label{eq: SR_phi_rho} 
    \end{align}
    where we sum over gluonic polarizations $\alpha=L, T$ and abbreviated $\int_{\mbf p}  = \int \mathrm{d}^2p/(2\pi)^2$. The approximation in the second line for each quantity rewrites these contributions employing the generalized fluctuation-dissipation relation in \eq \eqref{eq: fluct-diss-rel}. For the evaluation of the integrals, we use the statistical correlators $\langle EE \rangle (t, \omega, p)$ in momentum space computed in \cite{Boguslavski:2021buh} and revisited in \se \ref{subsec:prop_sectral_func}. The simulation setup in that work was similar to our study which is outlined in \se \ref{subsec:k_computation}. 

    Compared to the spectral functions, the statistical correlation functions of chromo-electric fields are numerically advantageous as the computation of spectral functions is known to become unstable over time. These instabilities restrict the accessible $\Delta t$-window and, thereby, the resolution in the frequency space. The statistical correlation function displays no such issues. Nevertheless, our results lead to important consequences for the spectral functions as \eqs \eqref{eq: SR_rho} and \eqref{eq: SR_phi_rho} connect our results directly to the gauge-fixed spectral functions.

    In the numerical reconstruction procedure of the heavy-quark diffusion coefficient following \eqs\eqref{eq: SR_EE} and \eqref{eq: SR_phi}, we cut off the momentum integrals at $p=7Q$ to reduce the computational cost of the extraction of the correlators at high momenta. This is justified since the weight of the equal-time correlation $\langle EE \rangle_\alpha(t,t,p)$ for $p > 7Q$ is close to zero as will be shown below for the Glasma-like theory in \fig\ref{fig: equal-time_corrs}, which suppresses such high-momentum excitations in the calculation of $\kappa$ and $\kappa_z$. The momentum integration range is then split up into three intervals where different techniques are employed to treat the integral. We discuss each interval in the following paragraphs.

    For small momenta, we numerically integrate the raw simulation data from \cite{Boguslavski:2021buh} up to a momentum $p^*\approx Q/2$. We ensured the robustness of our method by checking that the results are insensitive to small variations of $p^* > m_D$. 

    At larger momenta $ p^* \leq p < 2Q$ we combine the generalized fluctuation-dissipation relation and the characterization of the normalized spectral function in \eq\eqref{eq: SF_fit} to substitute the data with fitted analytical expressions. In the case of the transverse and scalar statistical correlators, they are dominated by their excitation peak and thus can be fitted by a (modified) Gaussian function $h_G^{\mrm{mod}}$ in \eq \eqref{eq:Gaussfit_mod}.%
    \footnote{
        Since we consider $p \geq p^* > m_D $, it is not necessary to include an additional transport peak $h^{\mrm{TP}}_\mrm{G}$ at $\omega=0$ in the fits as done in \figs\ref{fig: trans_comp_SF_Corrs} and \ref{fig: long_comp_SF_Corrs}, and the Landau damping contribution can be neglected as it is strongly suppressed.
    }
    For the longitudinal correlator at $p^* \leq p < 2Q$, the excitation peak is suppressed and it is well described by the Landau cut $\rho_L^\mathrm{Landau}$ (see \fig \ref{fig: long_comp_SF_Corrs}). We note that the Landau cut contribution is calculated as an analytical function of momentum and frequency where we used $m_\mathrm{HTL}=m_D$ from \eq \eqref{eq: asympt_mass} for the respective theory. This allows us to perform the integration over frequencies in \eqs \eqref{eq: SR_rho} and \eqref{eq: SR_phi_rho} over the fitted/analytical functions to improve the numerical accuracy.

    For the largest considered momentum interval $2Q \leq p < 7Q$, the peak position of transverse and scalar excitations was shown to follow an approximate relativistic dispersion relation 
    \begin{align} \label{eq: relat_disp_rel}
        \omega_R \approx \omega_\mathrm{rel}(p) = \sqrt{\omega_\mathrm{pl}^2 + p^2}
    \end{align}
    to good accuracy while the longitudinal peak is, with $\langle EE \rangle_L (t,t,p) \approx 0$, too strongly suppressed to provide any relevant information \cite{Boguslavski:2021buh}. To accelerate and numerically improve our reconstruction procedure, we use an analytical form where we approximate the spectral function for higher momenta using a Gaussian peak \eqref{Gaussfit} with the dispersion relation \eqref{eq: relat_disp_rel}. For the plasmon frequency $\omega_\mathrm{pl}$ we adopt the numerically extracted values $\omega_\mathrm{pl}^\mathrm{2D}=0.12Q$ and $\omega_\mathrm{pl}^\mathrm{2D+sc}=0.20Q$ for the 2D and Glasma-like theories \cite{Boguslavski:2021buh}. There it was also found that for sufficiently high momenta $p \gg \omega_\mathrm{pl}$ the values for the damping widths at $Qt=500$ are $p$-independent with $\gamma^\mathrm{2D} = 0.086Q$ and $\gamma^\mathrm{2D+sc} = 0.094Q$ for both the transverse and scalar polarizations. 

\subsection{Numerical reconstruction: results} \label{subsec:numer_reconst_results}
    In \fig \ref{fig:kappa_vs_bounds} we demonstrate the results of our reconstruction of the gluonic/scalar coefficients $2\kappa$ and $\kappa_z$ compared to approximations of the heavy-quark diffusion coefficients $\kappa^{\lessgtr}$ in \eqs \eqref{eq: k_lt} and \eqref{eq: k_gt} and $\kappa^{\mrm{def}}$ in \eq \eqref{eq:kappa}. The reconstructed curves $\kappa_\mrm{rec}$ show good agreement with these quantities both at early times (insets) and during the later evolution (main panels).%
    \footnote{
        \label{ft:scalars_small_Qt}
        While at earlier times $Q \Delta t \lesssim 40$ the reconstructed scalar $\kappa_z$ coincides with the gauge-invariant computations, one observes that deviations become more pronounced at later times. We note that this is an artifact propagating from the original correlation functions extracted in \cite{Boguslavski:2021buh}. There, it was shown that the damping rate is very small for the unequal-time correlator of the scalar contribution at soft momenta, and hence, its Fourier transformation to frequency space was not very precise. Since we use these data sets in our integration procedure, this error affects our late-time reconstruction.
    } 
    From the reconstruction procedure, we find that the early-time behavior is governed by high momentum contributions, while at late times soft momenta, and in particular their low-frequency properties described by the longitudinal Landau cut and the novel transport peak, dominate the diffusion coefficient. 

    We emphasize that the agreement between $\kappa$ and $\kappa_{z}$ and their respective reconstructed counterparts is non-trivial since one compares integrated gauge-fixed correlation functions $\langle EE \rangle_\alpha(t, \omega, p)$ with independently extracted manifestly gauge-invariant observables $\kappa_{(z)}(t, \Delta t)$. This demonstrates that these gauge-invariant quantities carry information on the features of the underlying microscopic excitations. We will use this result in the following section to obtain evidence for the existence of important features in the excitation spectrum.

\section{Testing the influence of unequal-time correlators on \texorpdfstring{$\boldsymbol{\kappa (t, \Delta t)}$}{k}} \label{sec:properties_reconst}
    The evolution of the gauge-independent heavy-quark diffusion coefficients $\kappa(t, \Delta t)$ that we computed in \se \ref{subsec:k_computation} can be successfully reconstructed using unequal-time correlation functions as described in \se \ref{sec:spectral_reconst}. In the next step, we will manipulate the unusual properties of these gauge-fixed correlation functions that we discussed in \se \ref{subsec:prop_sectral_func} before using them as input to the spectral reconstruction procedure. This allows us to study their influence on the diffusion coefficient by comparing them to the original evolution. With this method, we reinforce the evidence for the existence of the novel transport peak in the transverse correlation functions, for the surprisingly broad excitation peaks in the gluonic sector and the relatively narrow peaks in the scalar sector of the underlying excitation spectrum at low momenta, and for an infrared enhancement of the scalar sector.

    \begin{figure}[t!]
        \centering
        \includegraphics[width=0.49\textwidth]{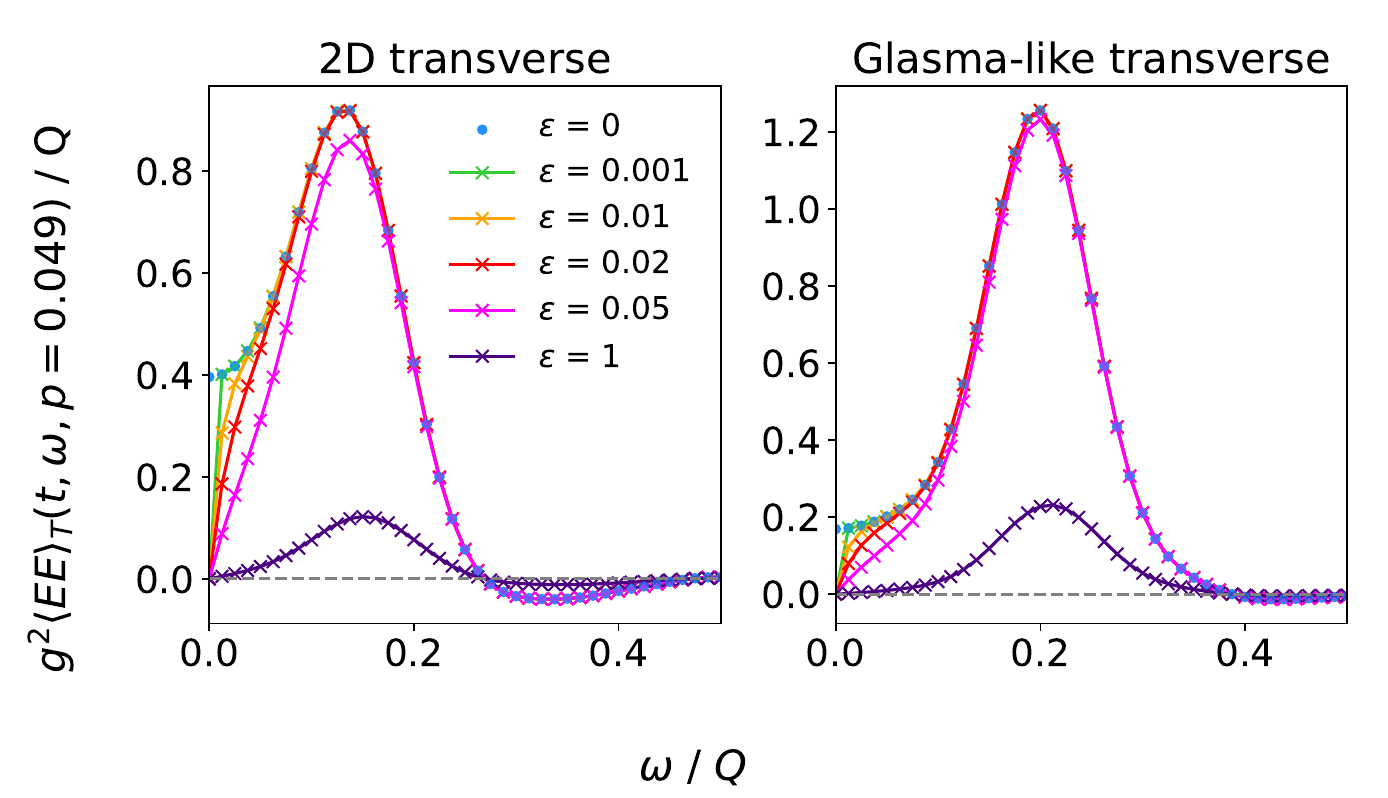}
        \caption{
            The transverse statistical correlator at $p=0.049 Q$ for the 2D \emph{(left)} and Glasma-like \emph{(right)} plasmas, modified by the factor in \eq\eqref{mod_factor}. For $\varepsilon > 0$, this suppresses the finite offset at $\omega = 0$ that is dominated by the transport peak.
        }
        \label{fig: cut_peaks}
    \end{figure}
    
    \begin{figure}[t!]
        \centering
        \includegraphics[width=0.4\textwidth]{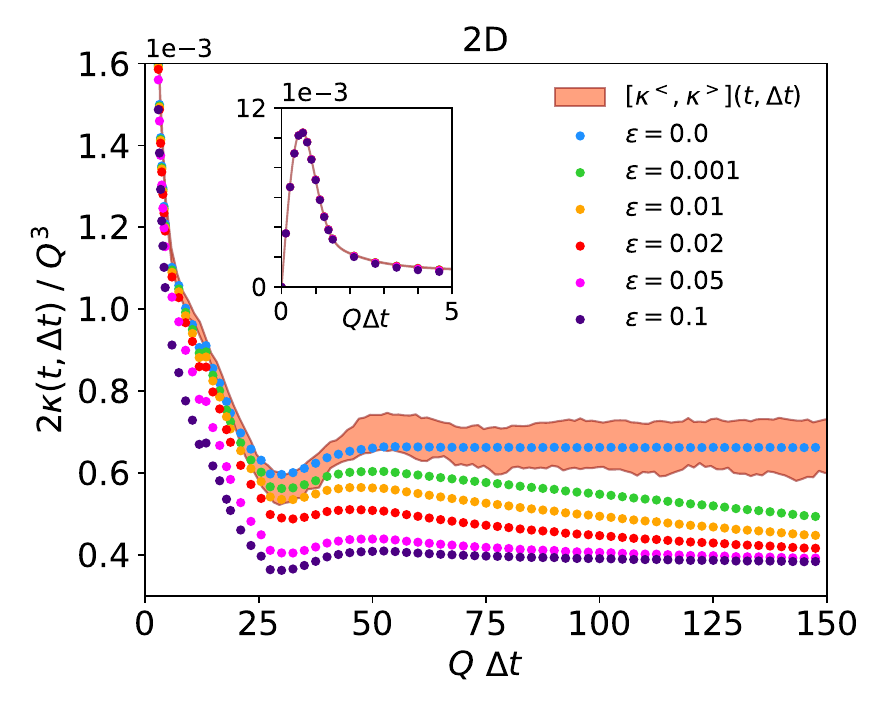}
        \includegraphics[width=0.4\textwidth]{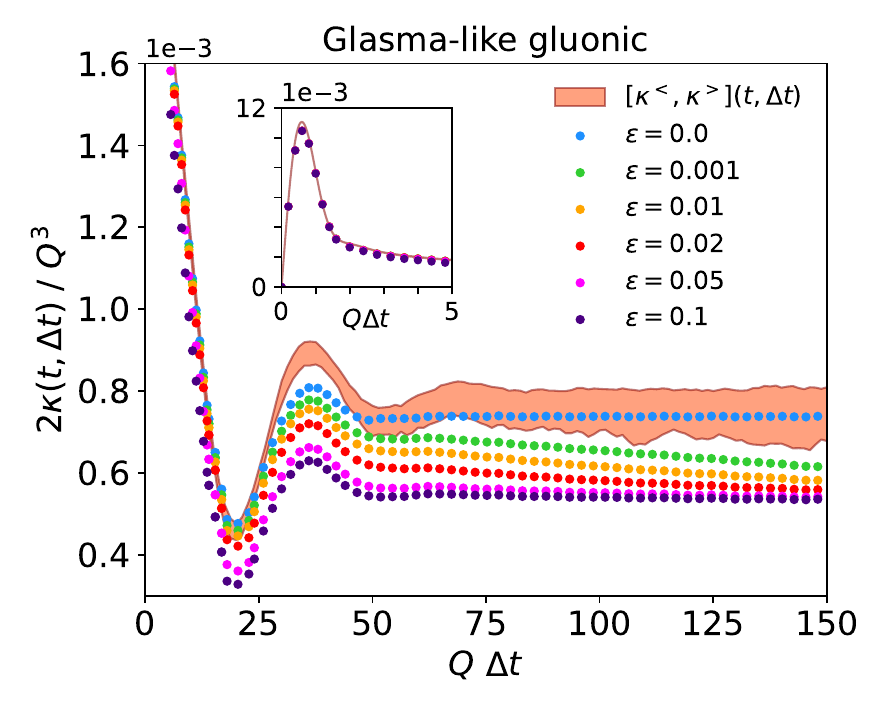}
        \caption{
            Comparison of the reconstructed heavy-quark diffusion coefficient for different suppressions $\varepsilon$ of the transverse correlator at low frequencies as given by \eq\eqref{mod_factor} and shown in \fig \ref{fig: cut_peaks}. The upper and lower panels show $2 \kappa$ of the 2D and Glasma-like plasmas as functions of time $\Delta t$. For comparison, the area between the unmodified coefficients $\kappa^\lessgtr$ is shown as an orange band. One observes significant deviations for any suppression $\varepsilon > 0$ at late $\Delta t$ indicating the crucial contributions of the transverse correlators at $\omega=0$ to $\kappa$, while the coefficient remains unchanged for small $\Delta t$ (\emph{insets}). \label{fig: kappa_eps}
        }
        \label{fig: trans_modulation}
    \end{figure}

\subsection{Evidence of a transport peak in \texorpdfstring{$\boldsymbol{\langle EE \rangle_T}$}{}}\label{subsec:low_frequ_behavior}
    In \se \ref{subsec:prop_sectral_func}, we have revisited that gluonic correlators have an offset at $\omega = 0$ \cite{Boguslavski:2021buh} and found that its accurate description requires the inclusion of a transport peak in $\langle EE \rangle_T$ and $\langle EE \rangle_L$ for small momenta $p \lesssim m_D$. Here we investigate the impact of the offset on the evolution of $\kappa$. 

    For this, we suppress the low-frequency behavior of transverse correlators using an exponential cutoff
    \begin{equation} \label{mod_factor}
        \langle EE \rangle_T (t, \omega, p) \mapsto (1 - e^{-\omega / \varepsilon}) \langle EE \rangle_T (t, \omega, p)
    \end{equation}
    with a suppression rate $\varepsilon$. The modified correlation functions are shown in \fig \ref{fig: cut_peaks} for $\varepsilon=0$ (no suppression) up to $\varepsilon=0.1$ (strong suppression) at $p=0.049Q$. We emphasize that for sufficiently large frequencies $\omega$, the curves remain effectively unchanged for the tested $\varepsilon$ range while at $\omega=0$, they are suppressed to $0$. The smallest modifying value $\varepsilon=0.001$ numerically suppresses only a single data point at $\omega=0$. For $\varepsilon=0$ we use the original correlation function.

    In \fig \ref{fig: trans_modulation}, we demonstrate the sensitivity of the heavy-quark diffusion coefficient to $\varepsilon$, i.e., to the suppression of the offset in $\langle EE \rangle_T$ (and thus in $\dot{\rho}_T$ due to \eq \eqref{eq: fluct-diss-rel}), by computing $\kappa$ in \eq \eqref{eq: SR_EE} using the modified correlations. We find that for all $\varepsilon > 0$, the coefficient decreases at sufficiently late times $\Delta t$, leaving the (systematic) uncertainty band for $\kappa$. Even for the weakest offset suppression $\varepsilon=0.001$ where only the data point at $\omega = 0$ is omitted, we observe a large influence on $\kappa$ at late $\Delta t$ (green curve). For all larger suppressions $\varepsilon \geq 0.1$, we observe that the reconstructed coefficient forms a plateau after $Q\Delta t \gtrsim 60$ without changing any further. Consequently, the contribution of the transverse correlation function to the late-time coefficient $\kappa_\infty$ is now suppressed entirely, and the evolution is completely determined by the unchanged longitudinal correlation function $\langle EE \rangle_L$ in \eq \eqref{eq: SF_fit}. Interestingly, the resulting values of our modified $\kappa_\infty$ are still larger than the perturbative expectation $\kappa_\infty^{\mathrm{pert}}$ from \se \ref{subsec:kappa_infty_pert} that we showed in \fig \ref{fig:kappa_inf}. Since the perturbative values originate solely from longitudinal Landau damping, our results indicate that apart from the transverse correlator, also an additional $\omega=0$ offset of the longitudinal correlator at low momenta contributes and is necessary to accurately describe $\kappa_\infty$.

    In contrast, at early $\Delta t$, the reconstructed curves are unchanged by the low-frequency suppression \eqref{mod_factor}, as shown in the insets of \fig \ref{fig: kappa_eps}. This consistency check confirms that the early-time behavior of $\kappa$ is governed by large frequencies at high momenta, which are unaffected by the suppression of the correlators at $\omega \approx 0$.

    Our analysis demonstrates that the $\omega=0$ offset of the transverse (and longitudinal) correlation function, and thus the new transport peak, is not an artifact of the gauge fixing procedure but rather an important physical feature as it strongly influences the transport properties of the plasma. In particular, by comparing the numerically extracted values from \fig \ref{fig: kappa_eps}, we find that the heavy-quark diffusion coefficient $\kappa_\infty$ increases by about $70 \%$ for the 2D plasma and by about $50 \%$ for the Glasma-like theory when the $\omega=0$ offset of the transverse correlator is taken into account. This shows that the non-perturbative offset leads to significantly larger values than what is naively expected when solely considering the Landau cut contribution in the longitudinal sector. Due to the gauge invariance of $\kappa(t, \Delta t)$, our results thus provide gauge-invariant evidence of the existence of the transport peak.

\subsection{Implications from broad and narrow excitations} \label{subsec:width}

    Another property of interest is the observed broad peak width (large damping rate) $\gamma_\alpha$ of the correlation functions in the gluonic sector, as well as the narrower width in the scalar sector \cite{Boguslavski:2021buh}. The interpretation of the broad peaks is that soft gluonic excitations for $p \lesssim m_D$ are too short-lived to form sufficiently stable quasiparticles that can participate in scattering processes. In contrast, the narrow scalar excitations imply a valid quasiparticle picture.

    We study the physical relevance of these features by analyzing the impact they have on $\kappa$. For this purpose, we fit our transverse and scalar correlators as modified Gaussian curves $h_\mrm{G}^\mrm{mod}$ in \eq \eqref{eq:Gaussfit_mod} for all $p \geq 0$ and manually vary $\gamma_\alpha$ before repeating the reconstruction procedure of \se \ref{sec:spectral_reconst}. To preserve the distinct offset of the correlation functions at low momenta, the fitting procedure occurs in three steps: 
    \begin{enumerate}
        \item We fit the correlators as Gaussian curves for some $\omega > 0$ to capture the shape of the dominant peak correctly.
        \item We subtract this fitted Gaussian curve from the correlator to unveil the transport peak's contribution around $\omega=0$ and fit it with a second Gaussian curve for small frequencies.
        \item We add both fitted contributions in \eqref{eq:Gaussfit_mod} to accurately describe the shape of the statistical correlation function at low momenta for the entire considered frequency range.
    \end{enumerate}

\begin{figure}[t!]
        \centering
        \includegraphics[width=0.34\textwidth]{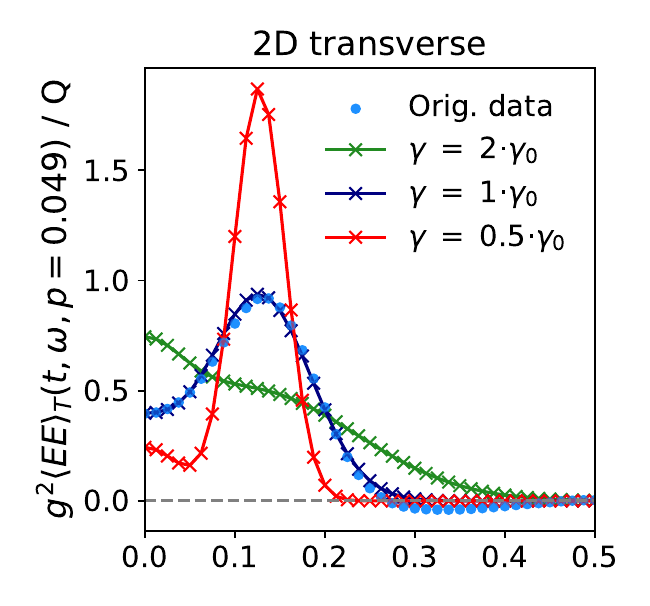}
        \includegraphics[width=0.34\textwidth]{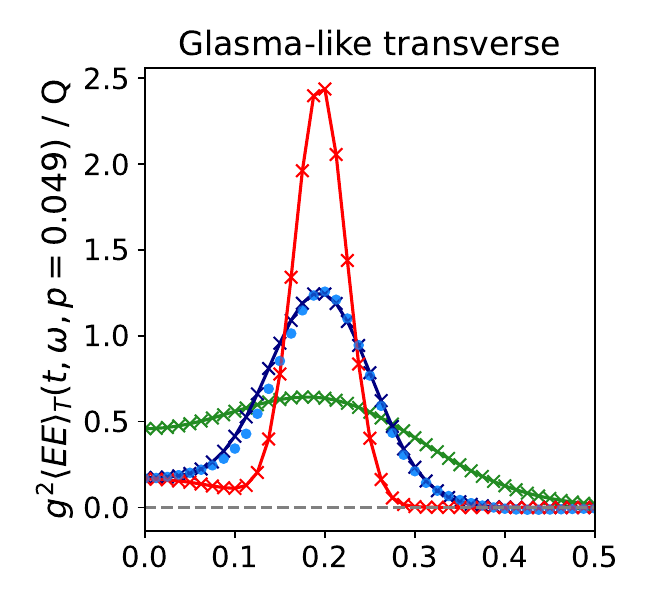}
        \includegraphics[width=0.34\textwidth]{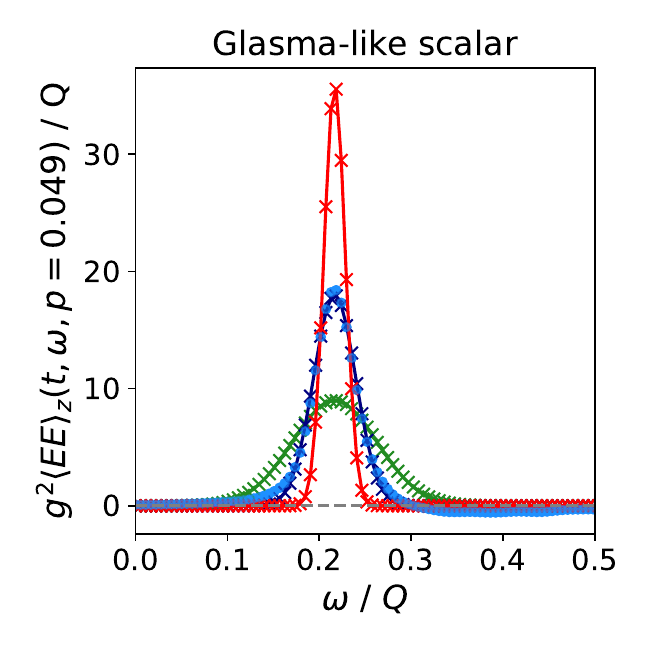}
        \caption{
            The transverse statistical correlator at $p=0.049 Q$ for the 2D ({\em top}) and the Glasma-like theory ({\em center}), as well as the scalar statistical correlation function of the Glasma-like plasma ({\em bottom}) as functions of frequency $\omega$. In addition, we show their modified Gaussian fits from \eq \eqref{eq:Gaussfit_mod} denoted as $\gamma = 1\cdot\gamma_0$ as well as excitations where we have doubled or halved the peak width of the fitted curves manually.
        }
        \label{fig: PW_peaks}
    \end{figure}

    The modified fitted correlators at $p=0.049Q$ are shown in \fig \ref{fig: PW_peaks} as functions of frequency for different peak widths $\gamma$ of the dominant Gaussian excitation $h_\mrm{Gauss}$ with the transport peak remaining unchanged throughout the procedure. Here $\gamma_0$ corresponds to the fitted peak width to the original data that is also shown for comparison. One observes that the modified Gaussian fit accurately reproduces the behavior of the data, as can be seen for the $\gamma = 1 \cdot  \gamma_0$ curve. Similarly, the fit reproduces the correlators in \figs \ref{fig: trans_comp_SF_Corrs} and \ref{fig: long_comp_SF_Corrs} where the new transport peak is explicitly shown for the normalized correlation functions. In contrast, by varying $\gamma$ in \fig \ref{fig: PW_peaks}, we observe significantly broader ($\gamma = 2\gamma_0$) and narrower ($\gamma = 0.5\gamma_0$) excitation peaks than the original peak. 

    \begin{figure}[t!]
        \centering
        \includegraphics[width=0.4\textwidth]{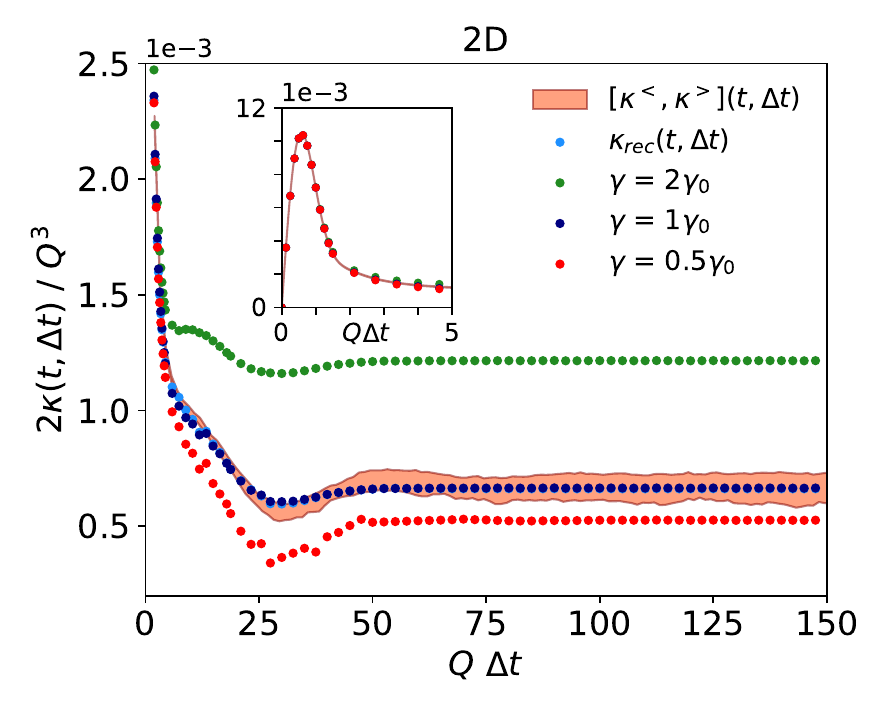}
        \includegraphics[width=0.4\textwidth]{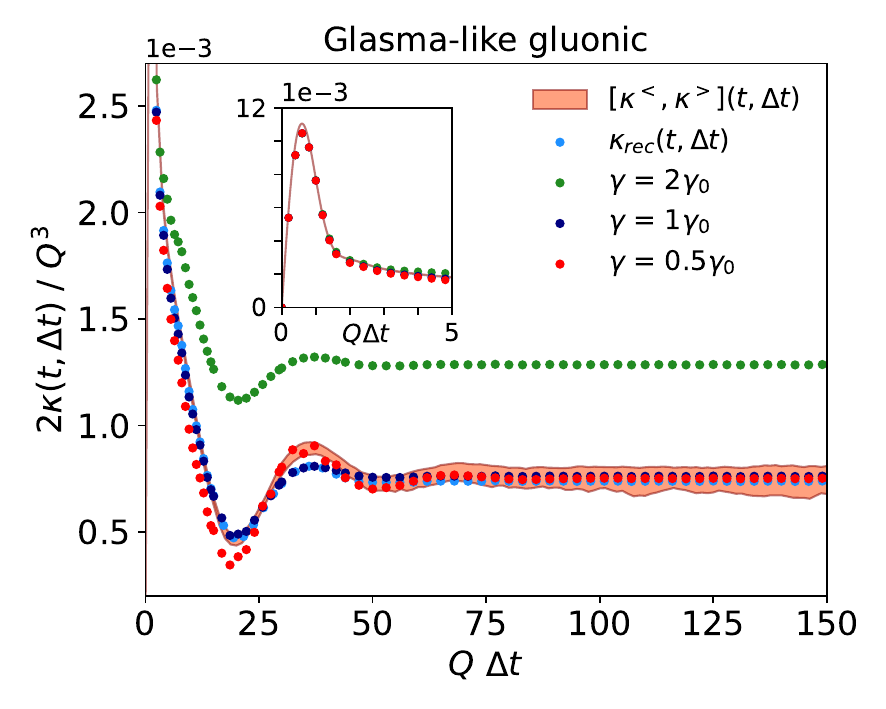}
        \includegraphics[width=0.4\textwidth]{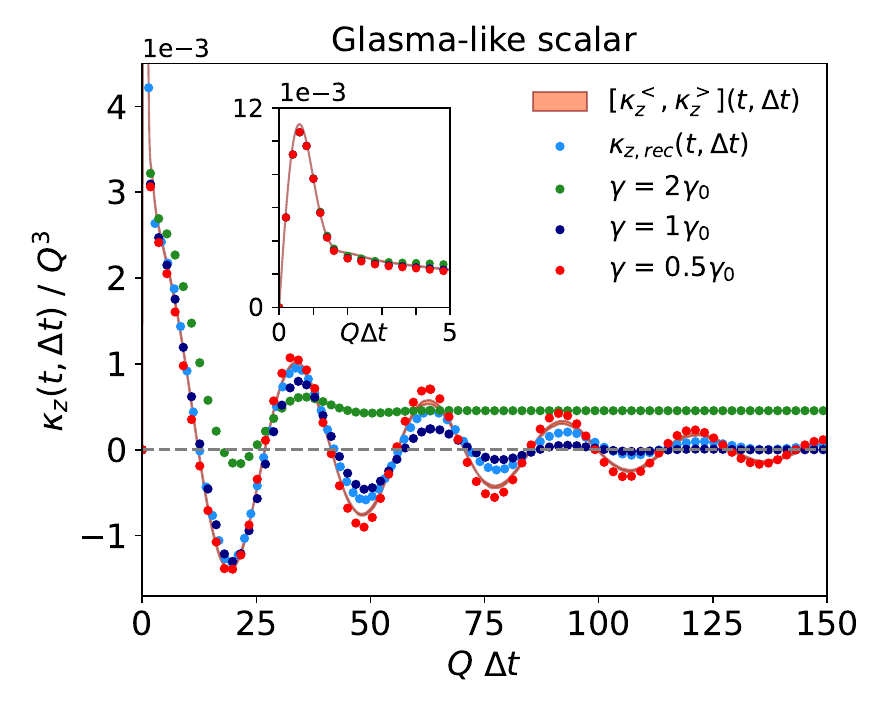}
        \caption{
            The reconstructed heavy-quark diffusion coefficients $2\kappa$ of the 2D ({\em top}) and the Glasma-like plasmas ({\em center}) and $\kappa_z$ ({\em bottom}). As in \fig \ref{fig: PW_peaks} we modify the damping rates $\gamma$ of the underlying excitation peaks of $\langle EE \rangle_T$ or $\langle EE \rangle_z$. They are compared to the original curves from \fig \ref{fig:kappa_vs_bounds}. In both systems, a strong sensitivity of the late $\Delta t$ behavior is observed upon varying $\gamma$, while the coefficient remains insensitive at early $\Delta t$ (\emph{insets}).
            \label{fig: kappa_peakwidth}
        }
    \end{figure}

    In \fig \ref{fig: kappa_peakwidth}, we demonstrate that the reconstructed $\kappa$ and $\kappa_z$ are sensitive to such variations of the damping rates for both theories. In the case of the gluonic $\kappa$ shown in the upper and central panels for the 2D and Glasma-like systems, we observe that broader excitation peaks increase the magnitude $\kappa_\infty$ of the diffusion coefficient at late $\Delta t$ while a narrower peak generally leads to a reduction of the value. These shifts can be understood as a consequence of the offset in the correlation functions at low momenta and $\omega = 0$ in \fig \ref{fig: PW_peaks} that is affected by the modified width. For the 2D theory, changing the width increases or suppresses the $\omega=0$ offset, influencing the value of $\kappa_\infty$ directly. For the transverse correlator of the Glasma-like theory, a larger $\gamma$ increases the offset and thus $\kappa_\infty$. However, a reduced peak width leaves the $\omega=0$ offset in the $\langle EE \rangle_T$ correlator unaffected at small momenta due to the (unmodified) transport peak that dominates the offset in this case. This results in a nearly unchanged $\kappa_\infty$ and corresponds to another (indirect) indication of the existence of the new transport peak. 

    \begin{figure}[t!]
        \centering
        \includegraphics[width=0.35\textwidth]{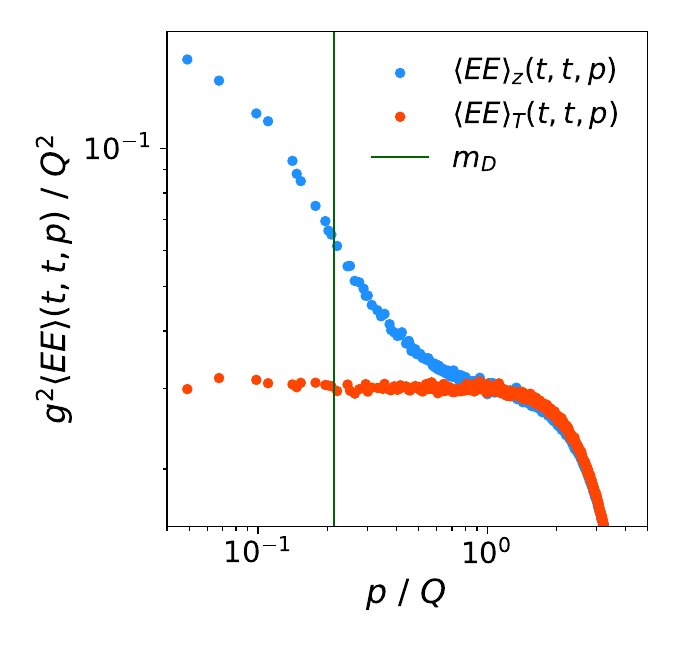}
        \includegraphics[width=0.35\textwidth]{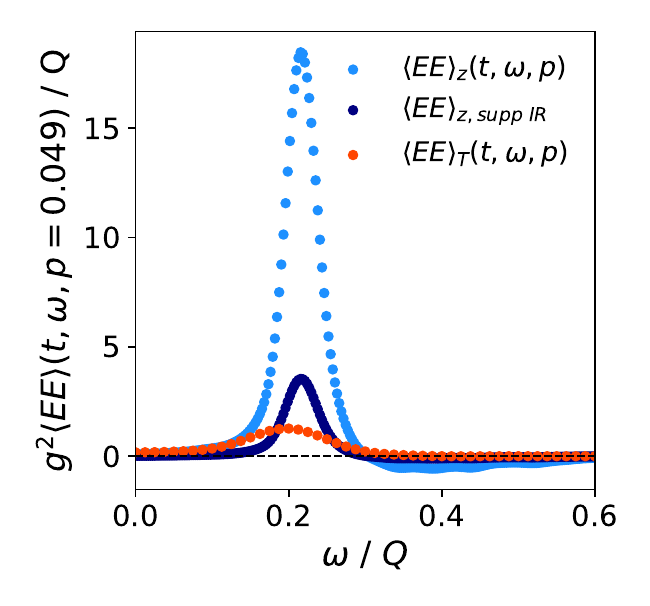} 
        \caption{
            {\em Top panel:} The equal-time correlation functions $\langle EE \rangle_z$ and $\langle EE \rangle_T$ as functions of momentum $p$. While agreeing at high momenta, the scalar correlator is enhanced at soft momenta. We have marked the value of the Debye mass with a green vertical line for comparison.
            {\em Bottom panel:} Statistical correlation functions at $p = 0.049 Q$ as functions of frequency $\omega$. We also show the scalar correlator $\langle EE \rangle_\mrm{z,\, supp\,IR}$ whose magnitude at low momenta has been rescaled according to \eq \eqref{eq:normfactor} using the equal-time correlators of the upper panel. 
        }
        \label{fig: equal-time_corrs}
    \end{figure}

    \begin{figure} [t!]
        \centering
        \includegraphics[width=0.45\textwidth]{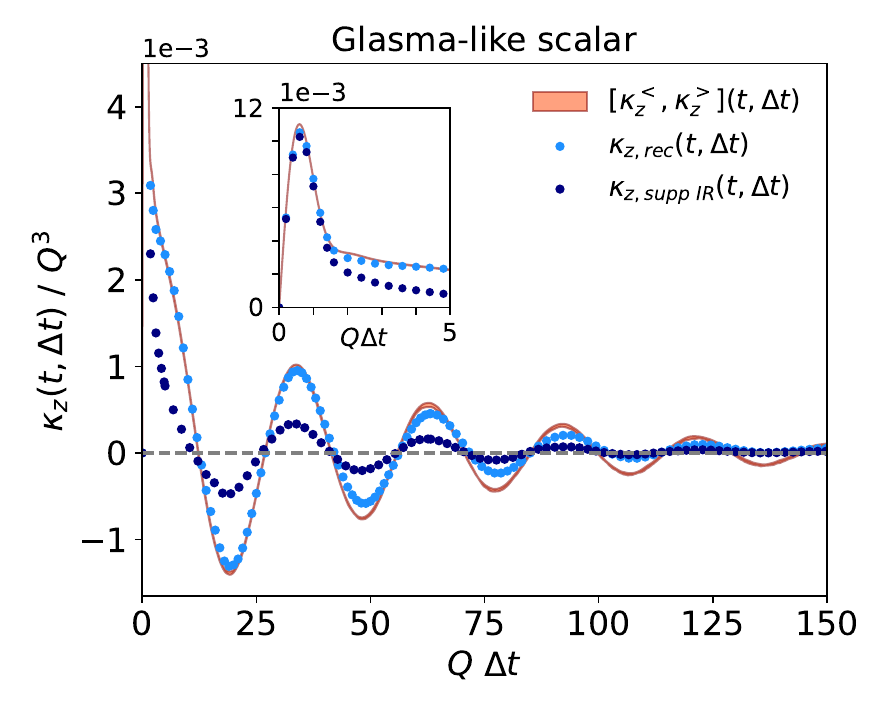}
        \caption{
            The original diffusion coefficient of the scalar sector $\kappa_z$ and the modified $\kappa_\mrm{z,\, supp\,IR}(t, \Delta t)$ as functions of $\Delta t$. In the latter, low-momentum excitations have been suppressed according to \eq \eqref{eq:normfactor}, which results in its smaller oscillation amplitude. $\kappa_{z, \mathrm{supp~IR}}$ compared to $\kappa_{z, \mrm{rec}}$. The inset shows early times where deviations occur already at $Q \Delta t \gtrsim 1$.
        }
        \label{fig: IR_enhancement}
    \end{figure}

    A more substantial effect appears in the scalar sector $\kappa_z$, where the excitation peaks at low momenta are narrow similar to the narrow peaks in 3+1D \cite{Boguslavski:2018beu}. An increase in the peak width at low momenta generates a noticeable offset at $\omega=0$ and leads to an evolution of $\kappa_z$ that closely resembles $2\kappa$ of the gluonic sector. This includes a shift of the late $\Delta t$ values upward with strongly damped late-time oscillations. Reducing the damping rate of the already narrow scalar excitation peaks effectively improves the agreement with the original $\kappa_z$. This indicates that we may have overestimated the width of the scalar excitations at low momenta, which could be a consequence of the Fourier transform in \cite{Boguslavski:2021buh} as mentioned in footnote \ref{ft:scalars_small_Qt}. We conclude that the distinct oscillations around zero, and thus the non-diffusive character of $\kappa_z$, seem to originate from the narrow quasi-particle-like excitations of the scalar sector.

    We show in the insets of \fig \ref{fig: kappa_peakwidth} that the early $Q \Delta t \lesssim 5$ evolution of both $\kappa$ and $\kappa_z$ remains unchanged by our width modifications. In this time interval, the integrals are dominated by hard excitations $p \sim \Lambda$ whose excitation peaks have a much higher frequency than their peak width $p \sim \omega \gg \gamma$. Modifying $\gamma$ thus does not have a sizable impact on the evolution. This is consistent with our previous interpretation that the peak width only of soft modes $p \lesssim m_D$ yields a sizable impact on the evolution causing visible shifts of $\kappa_\infty$.

\subsection{Evidence of an infrared enhancement of the scalar sector} \label{subsec:IR_enh}
    
    Throughout this work, the scalar correlator $\kappa_z (t, \Delta t)$ in the Glasma-like theory has been found to exhibit damped oscillations approximately around zero as a function of $\Delta t$, which is apparent in the lower panel of \fig \ref{fig:kappa_vs_bounds}. We argued in the previous section \ref{subsec:width} that these oscillations are connected to the narrow scalar excitation peaks of $\langle EE \rangle_z (t, \omega, p\lesssim m_D)$. We show here that there is another exciting factor contributing to the evolution of the scalar heavy-quark diffusion coefficient: an excess of soft gluons in the equal-time correlation function $\langle EE \rangle_z(t, t, p)$ at low momenta as compared to $\langle EE \rangle_T(t, t, p)$, which was observed in \re \cite{Boguslavski:2019fsb}. This infrared (IR) enhancement is shown in the upper panel of \fig \ref{fig: equal-time_corrs} where we show both equal-time correlators as functions of momentum. While they coincide and rapidly fall off toward 0 for higher momenta, the scalar correlator is strongly enhanced at small momenta. We aim to demonstrate that it affects $\kappa_z(t, \Delta t)$ by repeating its spectral reconstruction, but this time with a suppressed equal-time correlator at low momenta.

    We accomplish this by introducing a momentum-dependent suppression factor to the scalar correlator in the frequency domain as input for the reconstruction procedure,
    \begin{equation} \label{eq:normfactor}
        \langle EE \rangle_\mrm{z,\, supp\,IR} = \frac{\langle EE \rangle_T (t, t, p)}{\langle EE \rangle_z(t, t, p)} \cdot  \langle EE \rangle_\mrm{z} (t, \omega, p).
    \end{equation}
    The suppression of low momenta is achieved by the fraction $\langle EE \rangle_T (t, t, p)/\langle EE \rangle_z(t, t, p)$ while it approaches unity at higher momenta $p \gtrsim Q$. In the lower panel of \fig \ref{fig: equal-time_corrs}, we show the suppressed scalar correlator \eqref{eq:normfactor} representatively for $p=0.049Q$ together with the unmodified frequency-space correlation functions of the scalar and transverse polarizations. One indeed observes a much smaller amplitude of the peak in $\langle EE \rangle_\mrm{z,\, supp\,IR}$, which we will use to compute $\kappa_\mrm{z,\, supp\,IR}(t, \Delta t)$. We note that this strategy is reminiscent of \re \cite{Boguslavski:2020tqz}. The suppression of an infrared enhancement relative to perturbative estimates has been demonstrated to have a sizable impact on the evolution of $\kappa$ in 3+1 dimensional gluonic plasmas, causing suppressed oscillations in $\Delta t$. 

    A similar effect occurs for the reconstructed $\kappa_\mrm{z,\, supp\,IR}(t, \Delta t)$. It is plotted together with the original $\kappa_\mrm{z, rec}(t, \Delta t)$ in \fig \ref{fig: IR_enhancement}. Similar to the results observed in 3+1D, once the IR enhancement is suppressed, one observes considerably smaller oscillations in $\kappa_\mrm{z,\, supp\,IR}$. The inset reveals significant deviations already after a short time $Q \Delta t \gtrsim 1$. Thus, suppressing the low-momentum distribution of scalar excitations causes a distinct evolution of $\kappa_\mrm{z,\, supp\,IR}$. This can be understood as indirect evidence of an abundance of low-momentum excitations in the scalar sector, which corresponds to a genuinely non-perturbative property of $\langle EE \rangle_z$.

\section{Consequences for heavy-ion collisions} \label{sec:Glasma}
    This section discusses some phenomenological implications of our observations for heavy-ion collisions. 

    The pre-equilibrium matter produced initially in ultra-relativistic heavy-ion collisions is often described by the McLerran-Venugopalan model \cite{Gelis:2012ri, McLerran:1993ni, McLerran:1993ka, McLerran:1994vd} that is based on the color-glass-condensate description and is known as the Glasma. It shares essential features with the Glasma-like (2D+sc) theory studied in the present work. In particular, due to an approximate boost invariance, the Glasma's classical action can be effectively split into a 2+1 dimensional gluonic and scalar sector similar to \eq \eqref{eq: 2Dscaction}. Because of saturation, it consists of highly occupied gluonic fields at sufficiently soft momenta $p \lesssim Q$. In particular, the Glasma is described by 2-dimensional color-charge sheets whose chromoelectric and -magnetic fields point in the $z$ direction. Instead, we model saturation by choosing the initial conditions for both gluonic and scalar excitations of the form \eqref{eq: init_distrib_func}, i.e., we set highly occupied distributions for $p \lesssim Q$. The most crucial difference is that these Glasma-like plasmas evolve in Minkowski spacetime and not in a Bjorken expanding spacetime as for the Glasma. As a consequence, energy density is conserved in our case while it decreases with time in the expanding theory leading to an overall dilution of the system.

    The heavy-quark diffusion coefficient has been studied extensively in \res \cite{Carrington:2020sww, Boguslavski:2020tqz, Khowal:2021zoo, Avramescu:2023qvv, Das:2015aga, Mrowczynski:2017kso, Ipp:2020mjc, Pandey:2023dzz} in the Glasma and similar models. We will focus on a comparison with the recent study \cite{Avramescu:2023qvv}. Very similar to that work, we find a pronounced peak at early times $Q \Delta t \lesssim 1$ of the evolution of $\kappa$ for the gluonic (their transverse) as well as of $\kappa_z$ for the scalar (their longitudinal) contributions, as we show in the insets of \fig \ref{fig:kappa_vs_bounds}. They indeed grow linearly in $\Delta t$ and the slope is determined by the energy density of the respective sector, as explained in \eq \eqref{eq:kappa_early} and in \re \cite{Boguslavski:2020tqz}. This implies that the heavy-quark momentum broadens as $\langle p^2 \rangle \propto \Delta t^2$ initially. In our case, we have roughly the same energy density in the gluonic and scalar contributions, and thus, this early-time evolution is quantitatively similar for $2\kappa$ and $\kappa_z$ (see insets of \fig \ref{fig:kappa_vs_bounds}). In contrast, in the Glasma one finds $\kappa_z > 2 \kappa$ initially, which may hint at a slightly larger energy density in the scalar sector than for the gluonic components. 

    After this initial evolution, the coefficients $2\kappa$ and $\kappa_z$ evolve in qualitatively different manners for $Q \Delta t \gtrsim 1$, both in our Glasma-like system as well as in the expanding Glasma. For the gluonic sector, we find in our case that $2 \kappa$ stays positive, exhibits strongly damped oscillations with a frequency close to the plasmon frequency $\omega_\mrm{pl}$, and quickly approaches a plateau. We have demonstrated that this is due to broad gluonic excitations while a transport peak at $\omega = 0$ leads to a higher value of the plateau $2\kappa_\infty(t)$ than expected from longitudinal Landau damping. In the Glasma, $2\kappa$ also stays positive but approaches zero at late times. This difference is likely due to the system's aforementioned expansion, which dilutes the fields over time. In contrast, we observe for the Glasma-like case that $\kappa_z$ can become negative and generally exhibits weakly damped oscillations around zero. 
    We have found evidence that these oscillations are a consequence of narrow excitation peaks at low momenta in the scalar correlator $\langle EE\rangle_z$ (see the lower panels of \figs \ref{fig: PW_peaks} and \ref{fig: kappa_peakwidth}). Since $\kappa_z$ in the Glasma exhibits similar oscillations around zero, the relevant excitations in its scalar sector may also be narrow without any offset at $\omega=0$. Moreover, we find that in the Glasma-like system, the oscillations are enhanced from an excess of low-momentum modes (see \figs \ref{fig: equal-time_corrs} and \ref{fig: IR_enhancement}). Whether a similar effect exists in the Glasma should be checked in the future, together with our expectations for the structure of the correlators. 

    In conclusion, we emphasize that, despite manifest differences in these theories, we can draw physical implications from our non-expanding systems for the Glasma. In particular, we can explain the qualitative difference between the evolution of $2\kappa$ and $\kappa_z$ that was reported in \cite{Avramescu:2023qvv}. In an ongoing study, we extract the spectral and statistical correlation functions of the Bjorken expanding Glasma, and our preliminary results confirm the existence of broad peaks as well as excitations at $\omega=0$ in the gluonic sector, similar to those discussed in this work. This suggests that these properties, including the transport peak, are fundamental properties of highly occupied gluonic plasmas in 2+1 dimensions for different spacetimes and initial conditions. Similar features have been reported even in classical thermal equilibrium in 2+1 dimensions \cite{Boguslavski:2021buh}. Moreover, the transport peak seems to be essential for the transport properties of the Glasma, as it may help to explain the surprisingly large values of the transport coefficients during the Glasma stage. A detailed analysis similar to our study that focuses on transport coefficients in the Glasma is left for the future.


\section{Conclusion} \label{sec:conclu}
    In this paper, we have studied heavy-quark diffusion coefficients in 2+1 dimensional and Glasma-like gluonic plasmas. We identified distinctive properties of these transport coefficients for different directions and explained their origins using the underlying excitation spectrum where we distinguished gluonic and scalar sectors. Remarkably, our analysis provides evidence for non-perturbative features, such as broad gluonic excitations and a novel transport peak, which may significantly impact the evolution of hard probes in the Glasma.

    In particular, we have studied two related systems with high occupation numbers: a 2+1 dimensional Yang-Mills theory and a Glasma-like plasma. The latter extends the 2+1 dimensional plasma by introducing an additional scalar field component $A_z$ in similarity to the Glasma in heavy-ion collisions. We computed the evolution of the heavy-quark diffusion coefficients, distinguishing between $\kappa(t, \Delta t)$ and, in the case of the Glasma-like plasma, $\kappa_z(t, \Delta t)$, which are related to the gluonic ($x$, $y$) components and $z$ component of the chromo-electric field correlator, respectively. 

    We find that the gluonic coefficient $\kappa$ shows a fast growth followed by a subsequent decrease for early $Q \Delta t \lesssim 1$, and exhibits strongly damped oscillations at later times $Q \Delta t \gtrsim 1$ that eventually reach a plateau $\kappa_\infty$ that is associated with diffusion. In contrast, after a similar initial peak, $\kappa_z$ oscillates around zero with time, indicating that the scalar sector is non-diffusive. These features are very robust: For the considered highly occupied systems in their universal self-similar regimes, the $\Delta t$ evolution of $\kappa$ and $\kappa_z$ also shows a self-similar evolution in $t$. Therefore, these quantities do not change structurally. We also pointed out qualitative similarities with the Bjorken expanding Glasma including an initial peak and a different late-$\Delta t$ time evolution of $\kappa$ and $\kappa_z$.

    We were able to accurately reproduce the gauge-invariant evolution of $\kappa$ and $\kappa_{z}$ using integrals over gauge-fixed correlation functions $\langle EE \rangle_\alpha(t, \omega, p)$ and compare them with multiple definitions of the same coefficient, demonstrating their consistency. By manipulating the correlation functions of the chromo-electric field and studying the impact on the transport coefficients, we have found gauge-invariant evidence for the existence of
    \begin{itemize}
        \item a transport peak at $\omega = 0$ in gluonic $\langle EE \rangle$ that, in addition to longitudinal Landau damping, significantly contributes to the late-time values of $\kappa$, 
        \item soft gluons being short-lived while soft scalar excitations have extended lifetimes, which are visible as broad and narrow excitation peaks in gluonic $\langle E E \rangle$ and scalar $\langle E E \rangle_z$ for $p \lesssim m_D$, 
        \item and a relative enhancement of soft scalar excitations $\langle E E \rangle_z(t,t,p)$ as compared to gluons.
    \end{itemize}
    These properties of the underlying excitations have important consequences for non-Abelian plasmas in (effectively) 2+1 dimensions and are a manifestation of the non-perturbative microscopic structure of the theories. 

    The similarities of the studied systems with the 2+1D Bjorken expanding Glasma may allow us to infer microscopic features of the early stages of heavy-ion collisions. In particular, the transport peak and broad excitations of the gluonic correlations in the Glasma-like setting may provide an explanation for the relatively strong heavy-quark diffusion in the Glasma phase while the narrow and enhanced scalar excitations may explain the negative values of $\kappa_z$. A direct study of these features and their impact on the Glasma is in progress. We emphasize that these properties may also influence other transport coefficients or observables and could, therefore, serve as phenomenological signatures of non-perturbative and pre-equilibrium dynamics.

    Our study thus demonstrates that such lower dimensional plasmas exhibit genuinely non-perturbative features that can be seen in gauge-fixed unequal-time correlation functions and gauge-invariant transport coefficients. Their characterization provides invaluable insights into the non-perturbative structure of the theory and can help to formulate effective perturbative schemes that can be applied to lower dimensional gauge plasmas where HTL perturbation theory and kinetic theory fail. For instance, the transport peak is a crucial new ingredient for the formulation of scattering processes in 2+1 dimensional non-Abelian plasmas. Together with the other extracted features, a kinetic theory could be constructed that is consistent with Glasma-like systems and could become an alternative to classical-statistical simulations. 

    Another exciting prospect would be to compute $\kappa(t, \Delta t)$ in the full quantum field theory. While conventional lattice gauge theory \cite{Brambilla:2020siz, Banerjee:2022gen, Brambilla:2022xbd, Altenkort:2023oms, Altenkort:2023eav, Laine:2009dd} can only extract its late-time value $\kappa_\infty = \lim_{\Delta t \to \infty} \kappa(t, \Delta t)$ in thermal equilibrium, the use of complex Langevin for real-time correlation functions \cite{Boguslavski:2022dee, Boguslavski:2023unu} may enable a new avenue toward extracting the entire time evolution from first principles.


\begin{acknowledgments}
    We are grateful to Aleksi Kurkela, Tuomas Lappi, David I.~M\"uller, and Jarkko Peuron for valuable discussions and collaboration on related projects. This research was funded in whole or in part by the Austrian Science Fund (FWF) [10.55776/P34455]. For open access purposes, the author has applied a CC BY public copyright license to any author-accepted manuscript version arising from this submission. The results in this paper have been achieved using the Vienna Scientific Cluster (VSC) under project 71444.
\end{acknowledgments}


\appendix
\section{HTL expressions} \label{app:HTL}
    Hard thermal loop (HTL) perturbation theory provides a framework to compute correlation functions at soft (external) momenta using hard modes in the evaluation of loop integrals in gauge theory \cite{Braaten:1989mz, Blaizot:2001nr}. While for 3+1D systems the theory is very successful in its predictions, it does not cover the leading medium modifications for soft modes in 2+1D theories due to soft momenta obtaining more weight in momentum integrals, leading to deviations between HTL predictions and numerical simulation results of such models \cite{Boguslavski:2019fsb, Boguslavski:2021buh}. Nevertheless, it is still useful to compare our results with HTL expressions, and in the following, we review those in \cite{Boguslavski:2021buh} relevant to this work.

    Within the (leading-order) HTL approximation, one can define the HTL spectral function for each polarization $\alpha = T, L, z$ as the sum of a Landau damping contribution for $\abs{\omega} < p$ and a quasiparticle part,
    \begin{align} \label{eq: HTL_SF}
        \rho_\alpha^\mathrm{HTL}(\omega, p) 
        &= \rho_\alpha^\mathrm{Landau}(\omega, p) \\ 
        &+ 2 \pi Z_\alpha (p) \left[ \delta (\omega - \omega_\alpha^\mathrm{HTL}(p)) - (\omega\mapsto-\omega) \right].\nonumber
    \end{align}
    Here, $\omega_\alpha^\mathrm{HTL}(p)$ corresponds to a polarization-specific dispersion relation, and $Z_\alpha(p)$ are the residues of quasiparticle excitations obtained from the retarded propagators within the HTL framework. The dispersion relations allow the determination of the dominant frequencies for given momenta. While they generally differ for different polarizations, they predict that the transverse and scalar dispersion relations coincide for large momenta, which we can confirm numerically.

    In contrast to the delta-like peaks in \eq \eqref{eq: HTL_SF}, the excitation peaks exhibit a substantial width in 2+1D plasmas. We substitute the Dirac delta functions with a more general function $h(\omega, p)$ to describe the peak shape more appropriately.In \cite{Boguslavski:2021buh} the shape was shown to be well described by a Gaussian curve in \eq \eqref{Gaussfit}, which we confirm in \se\ref{sec:spectral_reconst}. 
    
    The Landau damping contributions in \eq \eqref{eq: HTL_SF} read
    \begin{align}
            \rho_T^\mathrm{Landau} & = \frac{2}{m_D^2} \frac{x \cdot \sqrt{1-x^2} \theta(1-x^2)}{((1-x^2)(p/m_D)^2 + x^2)^2 + x^2(1-x^2)}, \label{eq: T_Landau} \\
            \rho_L^\mathrm{Landau} & = \frac{2}{x} \frac{m_D^2/\sqrt{1-x^2}\theta(1-x^2)}{(p^2+m_D^2)^2+x^2m_D^4/(1-x^2)}, \label{eq: L_Landau} \\
            \rho_z^\mathrm{Landau} & = 0, \label{eq: phi_Landau}
    \end{align}
    where $x = \omega / p$. In HTL, the Debye mass $m_D$ is connected to the asymptotic mass that enters the dispersion relation at high momenta in 2+1D theories as $m_D^2 = m_\mathrm{HTL}^2$ and we calculate it in a self-consistent way using \cite{Boguslavski:2019fsb}
    \begin{align}
        m^2_\mathrm{HTL} = d_\mathrm{pol} N_c \int \frac{\mathrm{d}^d p}{(2\pi)^d} \frac{g^2 f(t,p)}{\sqrt{m_\mathrm{HTL}^2 + p^2}}.\label{eq: asympt_mass}
    \end{align}
    It is further connected to the plasmon frequency via $m_D^2 = 2 \omega_\mathrm{pl}^2$, albeit the numerically extracted $\omega_\mathrm{pl}$ from the excitation spectrum does not quite follow this relation to $m_D$ \cite{Boguslavski:2021buh}. It also enters the sum rules satisfied by the dotted spectral functions,
    \begin{align}
        \dot{\rho}_T (t, \Delta t=0, p) & = \int_{-\infty}^\infty \frac{\mathrm{d}\omega}{2\pi} \omega \rho_T (t, \omega, p) = 1 \label{eq: T_sumrule} \\
        \dot{\rho}_L (t, \Delta t=0, p) & = \int_{-\infty}^\infty \frac{\mathrm{d}\omega}{2\pi} \omega \rho_L (t, \omega, p) = \frac{m_D^2}{p^2 + m_D^2} \label{eq: L_sumrule} \\
        \dot{\rho}_z (t, \Delta t=0, p) & = \int_{-\infty}^\infty \frac{\mathrm{d}\omega}{2\pi} \omega \rho_z (t, \omega, p) = 1\,, \label{eq: phi_sumrule}
    \end{align}
    which yield normalization factors in the generalized fluctuation-dissipation relation in \eq \eqref{eq: fluct-diss-rel} and consequently in the spectral reconstruction procedure of \ses \ref{sec:spectral_reconst} and \ref{sec:properties_reconst}.

    We emphasize that for the analytical fits and approximations employed in this work, the momentum sector $p \gtrsim m_D$ for the transversely and scalar polarized correlation and spectral functions is governed by the quasi-particle peaks while the longitudinal polarization is predominantly described by the Landau cut.

\section{Scaling behavior of fit parameters} \label{app:self_sim_parameters}
    \begin{figure*}[t!]
        \centering
        \includegraphics[width=0.75\textwidth]{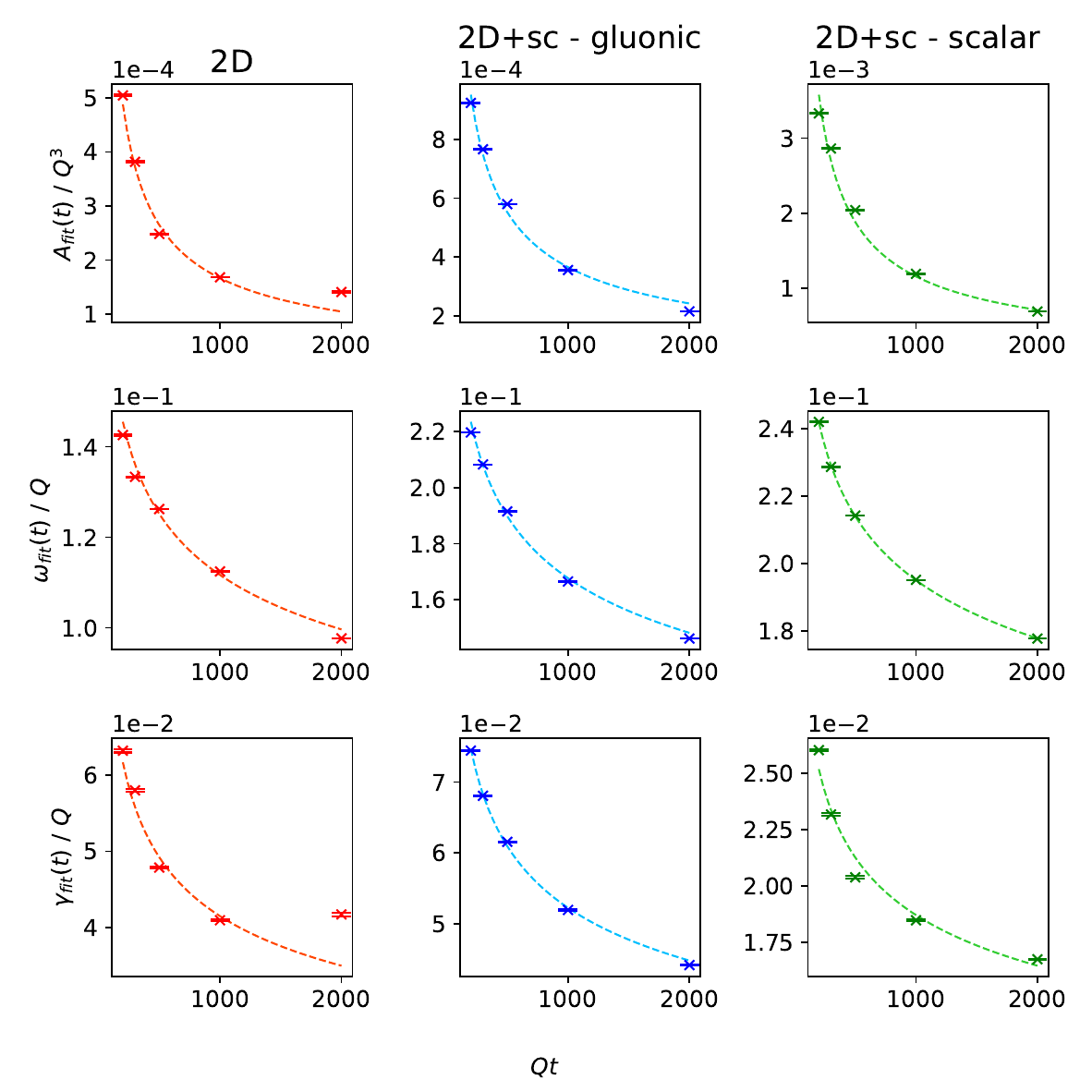}
        \caption{
            Scaling behaviors of the extracted fit parameters from $\kappa_\mathrm{fit}(t, \Delta t)$ in \eq \eqref{eq: damp_HO}. The rows show the fitted amplitude $A_\mathrm{fit}(t)$, frequency $\omega_\mathrm{fit}(t)$, and damping rate $\gamma_\mathrm{fit}(t)$ as functions of time $t$ for the gluonic and scalar sectors of the considered models. The dashed lines show curves fitted to the evolution of the parameters using power laws $\sim (Qt)^{-\sigma}$ in a least square fitting routine. Their respective scaling exponents are listed in \tab \ref{tab: fit_param}.
        }
        \label{fig:params_scaling}
    \end{figure*}
 
    In \se \ref{subsec:k_selfsim}, we discussed the self-similar evolution of the heavy-quark diffusion coefficient $\kappa$ with respect to $t$, as well as of its late-$\Delta t$ limit, $\kappa_\infty^\mathrm{fit}(t)$, which is shown in \fig\ref{fig:kappa_inf}. This plateau decreases approximately as $\kappa_\infty^\mathrm{fit}(t) \sim Q^3(Qt)^{-3/5}$ and is also compared to its perturbative estimates in that section. Here, we address the other fit parameters entering \eq \eqref{eq: damp_HO} in more detail.

    \begin{table}[b]
    \caption{
        \label{tab: fit_param} 
        Scaling exponents of the time-dependent heavy-quark diffusion coefficient $\kappa_{\mathrm{fit}}$'s fit parameters defined in \eq\eqref{eq: damp_HO}, determined by a least square fit. The offset and amplitude are denoted by $\kappa_\mathrm{fit}^\infty$ and $A_\mathrm{fit}$ while the frequency and decay rate of the harmonic oscillation are written as $\omega_\mathrm{fit}$ and  $\gamma_\mathrm{fit}$ respectively. The offset for the scalar sector in the 2D+sc setting is zero and, hence, does not exhibit a power law behavior.
    }
    \begin{ruledtabular}
        \begin{tabular}{ccccc}
            Scaling exponent & $\kappa^\infty_\mathrm{fit}$ & $A_\mathrm{fit}$ & $\omega_\mathrm{fit}$ & $\gamma_\mathrm{fit}$ \\
            \hline & \\[-0.75em]
            $\sigma^\mathrm{2D}$ & $0.58(1)$ & $0.67(4)$ & $0.16(1)$ & $0.25(4)$ \\[0.5em]
            $\sigma^\mathrm{2D+sc}_\mathrm{gluonic}$ & $0.56(1)$ & $0.59(4)$ & $0.18(1)$ & $0.22(1)$ \\[0.5em]
            $\sigma^\mathrm{2D+sc}_\mathrm{z}$ & $-$ & $0.70(3)$ & $0.130(1)$ & $0.18(2)$ \\[0.25em]
        \end{tabular}
    \end{ruledtabular}
    \end{table}
    These parameters are well described by a scaling law $\sim Q(Qt)^{-\sigma}$. We have summarized the resulting scaling exponents from a least square fitting procedure through our numerical data in \tab \ref{tab: fit_param} for the considered theories and both gluonic and scalar sectors. The extracted parameters are shown with the corresponding fitted power laws as functions of time in \fig\ref{fig:params_scaling}. While they generally agree well with the power law fits, slight deviations between the data and the curves are still visible, especially for the 2D theory. This is not surprising since the 2D correlators exhibit weaker initial oscillations and larger error bars than their gluonic 2D+sc counterparts as is visible in the upper panel of \fig \ref{fig:kappa_OV}. Approximating them as damped harmonic oscillators is, therefore, more difficult and prone to numerical inaccuracies, which is reflected in the 2D scaling curve fits not being as accurate as for the Glasma-like system. The top row in \fig\ref{fig:params_scaling} shows the amplitude $A_\mathrm{fit}$ of $\kappa_\mathrm{fit}(t, \Delta t)$ while the center and bottom rows show the frequency $\omega_\mathrm{fit}$ and decay rate $\gamma_\mathrm{fit}$.
    
    We emphasize that scaling exponents of the frequency and decay rate are quantitatively similar to $\beta = -1/5$ for the gluonic components. This could suggest the relations $\omega_\mathrm{fit} \sim \omega_\mathrm{pl} \sim (Qt)^{-1/5}$ and $\gamma_\mathrm{fit} \sim \omega_\mathrm{pl}$ and a particular importance of soft momentum modes $p \lesssim m_D$ for $\kappa(t, \Delta t)$. We note that the larger discrepancy for the 2D theory scaling exponents is likely caused by the ill-conditioned fitting procedure of $\kappa$ at different times as mentioned before. Especially for $\gamma_\mathrm{fit} (t)$ the consequences are significant as there are strong deviations from the power law scaling at late times as seen in the bottom left panel of \fig\ref{fig:params_scaling}, which may lead to a large systematic error of its scaling exponent.

    The final parameter in \eq \eqref{eq: damp_HO} is the phase shift $\phi_\mrm{fit}$. This quantity does not have a physical interpretation and serves to improve the quality of the overall fit and to deal with the different $\Delta t$ time windows of the initial peak in $\kappa(t, \Delta t)$. We fix it to the following values that we have obtained by fitting \eq \eqref{eq: damp_HO} at the earliest time $Qt=200$,
    \begin{align*}
        \phi^\mrm{2D}_\mrm{fit} = 1.12, \quad
        \phi^\mrm{2D+sc}_\mathrm{gl, fit} = 0.98, \quad
        \phi^\mathrm{2D+sc}_\mathrm{z, fit} = 1.01.
    \end{align*}


\bibliographystyle{JHEP-2modlong}
\bibliography{references}

\end{document}